\documentclass[fleqn,usenatbib]{mnras}

\usepackage{newtxtext,newtxmath}

\usepackage[T1]{fontenc}

\DeclareRobustCommand{\VAN}[3]{#2}
\let\VANthebibliography\thebibliography
\def\thebibliography{\DeclareRobustCommand{\VAN}[3]{##3}\VANthebibliography}

\usepackage{graphicx}	
\usepackage{amsmath}	
\usepackage{caption}
\usepackage[table]{xcolor}
\usepackage{multirow}

\title[Massive Compact Galaxies in IllustrisTNG]{The Origin of Massive Compact Galaxies: Lessons from IllustrisTNG}

\author[F. S. Lohmann et al.]{
Felipe S. Lohmann,$^{1}$\thanks{E-mail: felipeslohmann@gmail.com}
Allan Schnorr-Müller,$^{1}$
Marina Trevisan,$^{1}$
T. V. Ricci$^{2}$
and K. Slodkowski Clerici$^{1}$
\\
$^{1}$Universidade Federal do Rio Grande do Sul $-$ Departamento de Astronomia $-$ 91501-970, Porto Alegre-RS, Brazil\\
$^{2}$Universidade Federal da Fronteira Sul – Campus Cerro Largo – 97900-000, Cerro Largo-RS, Brazil\\
}

\date{Accepted XXX. Received YYY; in original form ZZZ}

\pubyear{2015}

\begin{document}
\label{firstpage}
\pagerange{\pageref{firstpage}--\pageref{lastpage}}
\maketitle

\begin{abstract}
We investigate the formation and evolution of z=0 massive compact galaxies (MCGs) in the IllustrisTNG cosmological simulation. We found that, as in observations, MCGs are mainly old (median age $\sim 10.8$ Gyr), have super-solar metallicities (median $\log Z/Z_{\odot}\sim0.35$) and are $\alpha$-enhanced (median $[\alpha/Fe]\sim0.25$). The age distribution extends to younger ages, however, and a few MCGs are as young as $\sim7$\,Gyr. In general, MCGs assemble their mass early and accrete low angular momentum gas, significantly increasing their mass while growing their size much slower. A small fraction of MCGs follow another evolutionary path, going through a compaction event, with their sizes shrinking by 40\% or more. The accretion of low angular momentum gas leads to enhanced SMBH growth, and MCGs reach the threshold SMBH mass of $\log M_\mathrm{BH}\sim10^{8.5}\,M_\odot$ - when kinetic AGN feedback kicks in and quenches the galaxy - earlier than non-compact galaxies. Comparing MCGs to a sample of median-sized quiescent galaxies matched in effective velocity dispersion, we find that their accretion histories are very different. 71\% of MCGs do not merge after quenching compared to 37\% of median-sized quiescent galaxies. Moreover, tracing these populations back in time, we find that at least a third of median-sized quiescent galaxies do not have a compact progenitor, underscoring that both dry mergers and progenitor bias effects are responsible for the differences in the kinematics and stellar population properties of MCGs and median-sized quiescent galaxies.
\end{abstract}




\section{Introduction}


Galaxies are typically divided into two large groups: star-forming galaxies, where stellar mass and star formation rate are strongly correlated, and quiescent galaxies, where the formation of new stars has been quenched or is residual \citep{brinchmann04, noeske07}. The first quiescent galaxies emerge by $z \sim 4-5$ \citep{straatman14,carnall23}, and by $z \sim 2$ they are already dominant at the high end of the galaxy mass function \citep{muzzin13,Ilbert13}.  These high redshift quiescent galaxies differ significantly from their z\,$\sim$\,0 counterparts: they are extremely compact, being smaller than local quiescent galaxies of the same mass by a factor of $3-5$ \citep{vanderwel14}. Considering that the number density of massive ($\sim 10^{11}\,M_\odot$) compact quiescent galaxies declines by a factor of $\sim 10$ from $z \sim 1.5$ to $z \sim 0.5$ \citep{vanderwel14}, the size evolution of the quiescent population must be driven, at least in part, by the growth of individual galaxies. In cosmological simulations massive compact quiescent galaxies grow by repeated dry minor mergers \citep{oser10,furlong15,wellons16}, a scenario which is supported by observations of the most massive ($\log M_\star/M_\odot \gtrsim 11$) local quiescent galaxies \citep{huang13,greene13,oh17}. At lower masses, however, observations suggest that a progenitor bias effect, due to larger star-forming galaxies becoming quiescent at later epochs \citep{carollo13}, plays a larger role \citep{fagioli16,faisst17,damjanov19}. 

The progenitors of massive compact quiescent galaxies are believed to be the rare compact star-forming galaxies \citep{barro13,barro17}. Observations show that massive compact star-forming galaxies have high star formation rates (SFR), of the order of hundreds to thousands of solar masses per year \citep{barro13}, have low gas-to-stellar mass ratios \citep{spilker16,barro17b,talia18,spilker19} and frequently host an active galactic nuclei \citep{barro13,kocevski17,aird22}. These suggest compact star-forming galaxies are experiencing a centrally concentrated burst of star formation triggered by massive gas inflows, which also feed the supermassive black hole, and they will soon quench. 
Two origins have been suggested for compact star-forming galaxies: I) they form in a compaction event \citep{dekel14}, when violent disk instabilities drive large amounts of gas to the center of an extended star-forming galaxy where it fuels a starburst \citep{zolotov15,lapiner23}; II) they descend from star-forming galaxies that were already born compact and which have increased their mass without significantly increasing their size \citep{vandokkum15,suess21}.

There exists a rare population of galaxies that are still compact in the present-day universe. This population shows a range of star formation histories \citep{papermanga, spiniello21}, being composed of a mix of young \citep{ferre12} and old passively evolving objects \citep{yildirim17}. In a recent work \citep[hereafter SM21]{papermanga}, we studied the kinematics and stellar populations properties of a sample of massive compact quiescent galaxies (MCGs) extracted from the Mapping Nearby Galaxies
at Apache Point Observatory (MaNGA) survey \citep{smee13, bundy15, drory15} and we found that MCGs are predominantly old, metal rich and $\alpha$-enhanced, in agreement with previous studies with smaller samples \citep{yildirim17,buitrago18,spiniello21}. They also show significant rotational support, suggesting they have been passively evolving since quenching. Compared to a sample of median-sized quiescent galaxies matched in central velocity dispersion, MCGs have similar luminosity-weighted ages within $0.5 r_{\rm e}$, but higher metallicities and $\alpha$ element abundances. Due to the limitations inherent to archaeological studies, a number of questions remained open. For example, we could not determine whether the differences in stellar population properties are mainly due to the differences in accretion history or due to MCGs and median-sized quiescent galaxies descending from different progenitor populations (i.e. progenitor bias). 

Further progress in answering these open questions can be achieved through the study of cosmological simulations, such as the IllustrisTNG\footnote{https://www.tng-project.org/} suite \citep{springel18, nelson18, pillepich18a, naiman18, marinacci18}.  TNG successfully reproduces numerous scaling relations of galaxies in the local universe, such as the fundamental plane \citep{lu20}, the mass\,--\,metallicity relation \citep{torrey19}, the M$_\textrm{BH}$\,--\,M$_\textrm{Bulge}$ relation \citep{weinberger18} and the star formation main sequence \citep{donnari19}, in addition to reproducing fundamental properties of the population of galaxies such as the distribution of galaxies in the color-magnitude diagram \citep{nelson18}, the luminosity function of galaxies \citep{pillepich18a}, the dark matter fraction \citep{lovell18} and the morphology distribution of galaxies \citep{rodriguez-gomez19}. 

The aim of the work is twofold: 1) to characterize the kinematics, stellar populations properties and environment of MCGs and of a control sample of median-sized quiescent galaxies at $z=0$ to explore to which degree the trends reported by SM21 are reproduced by IllustrisTNG; 2) to trace these populations back in time, with the goal of understanding how MCGs form and to determine if MCGs and control sample galaxies descend from the same star-forming population. 

This work is structured as follows: we describe our methodology in Section 2, where we give general details of the TNG100 simulation, followed by the description of our selection of compact galaxies and the methods to calculate their properties. In Section 3 we show our results and compare the properties of the simulated galaxies with observations, while leaving a detailed discussion to Section 4. Finally, we summarise our conclusions in Section 5.

\section{Methodology}
\label{sec:methodology}

\subsection{The IllustrisTNG Simulation}

The IllustrisTNG project \citep{springel18, nelson18, pillepich18a, naiman18, marinacci18} is a series of magnetohydrodynamical cosmological simulations, corresponding to a continuation of the well known Illustris project \citep{vogelsberger13, vogelsberger14a, vogelsberger14b, genel14, nelson15}. 
As its predecessor, TNG uses the quasi-Lagrangian moving-mesh code \texttt{AREPO} \citep{springel10}, but covering larger volumes, at higher resolution and featuring an updated galaxy formation model, called the Fiducial TNG Model \citep{weinberger17, pillepich18b}. 
The TNG suite includes three primary runs with different resolution levels, called TNG50, TNG100, and TNG300. 
Considering MCGs are rare objects, we need both a high spatial resolution and a large simulated volume. With this in mind, we chose to primarily use the intermediate resolution run, TNG100, as it provides the best compromise between spatial resolution and simulated volume. From the available runs of TNG100, we chose the highest resolution one, referred to as TNG100-1.
This run features a comoving box volume of $110.7^3$ cMpc$^3$ with $1820^3$ dark matter and gas particles, with a starting redshift of $z=127$ and 99 publicly available snapshots separated by $\sim 150$ Myr. 
The cosmology adopted in the simulation is consistent with the Planck  measurements \citep{planck16}, with a total matter density $\Omega_m = 0.3089$ and reduced Hubble constant $h = 0.6774$ \,km\,s$^{-1}$\,Mpc$^{-1}$.

In order to identify halos and subhalos (i.e. simulated galaxy clusters and individual galaxies), TNG employs the \texttt{FoF} and \texttt{SUBFIND} \citep{springel01, dolag09} substructure identification algorithms, respectively. 
In order to track subhalos across snapshots, TNG uses the \texttt{SubLink} method \citep{RG15} to construct merger trees, identifying all subhalos that merged together in the past to form the observed subhalo. 
The main progenitor of each subhalo is defined as the one with the most massive history behind it, and thus the main progenitor branch is obtained connecting all main progenitors throughout all snapshots. 

The merger tree structure is also used to identify merger events between subhalos.
In this work, we consider as major mergers those where the mass ratio is greater than $1/4$, while those with mass ratio between $1/4$ and $1/10$ are considered minor mergers.

\subsection{Sample selection}

\begin{figure*}
    \centering
    \includegraphics[scale=0.55]{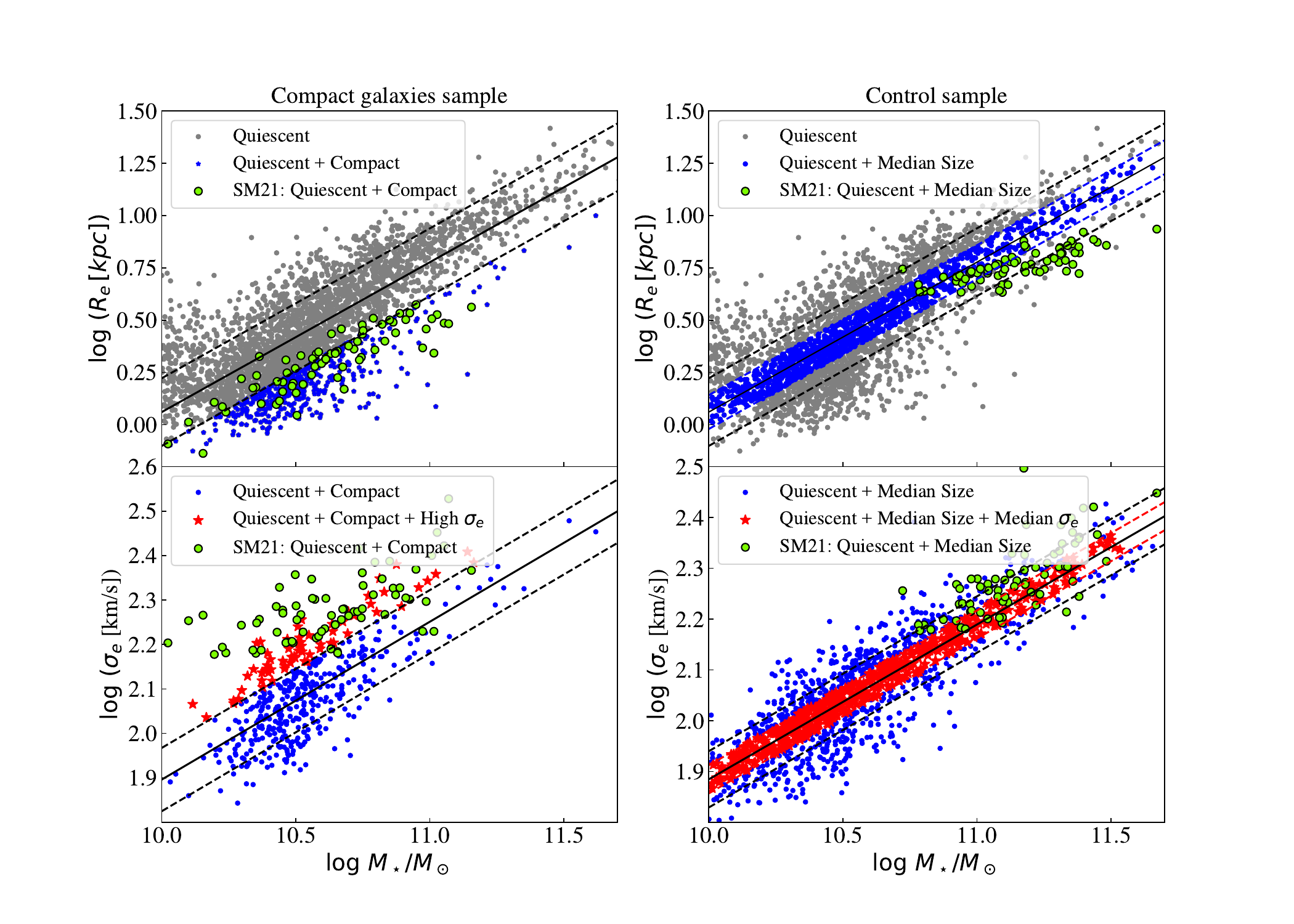}
    \caption{The two steps of our sample selection method, both for the massive compact galaxies sample (left column) and for the control sample (right column). The top panels represent the first step in the selection, and show the mass-size relation of quiescent galaxies in TNG100 as a solid line and the $1\sigma$ interval as dashed lines. The selected galaxies in this step is shown as blue symbols for each sample, following the respective size criteria. The bottom panels correspond to the second step in the selection, where the relation between the stellar mass and the velocity dispersion inside one half-mass radius $\sigma_e$ of the previously selected galaxies is used. The galaxies selected in this step following the respective $\sigma_e$ criteria are shown as red stars. The green circles show the locations of the MaNGA quiescent galaxies studied by SM21.}
    \label{fig:sample_selec}
\end{figure*}

Ideally, we would like to select our sample of MCGs applying the same criteria as SM21. However, SM21 relied on the size-velocity dispersion relation, which is not well reproduced in TNG100, likely due to the simulation underpredicting the velocity dispersion of galaxies \citep{lu_cappellari20, wang20}. For this reason, we chose instead to select MCGs according to two criteria: being outliers on the mass-size relation (as the size-mass relation is well reproduced by TNG, \citealt{genel18}) and having higher than average velocity dispersion for their stellar mass (so that galaxies in our sample have small sizes and high velocity dispersion as the MCGs in SM21).

\subsubsection{Massive Compact Galaxies}

We start by selecting only galaxies that are quiescent at $z=0$ in TNG100-1.  We considered as quiescent those galaxies with specific star formation below $10^{-11}$ yr$^{-1}$. Next, we select outliers on the mass-size relation. We obtained the mass-size relation of quiescent galaxies by performing a least squares linear fit to the stellar mass and half-mass radius. We proceed to select galaxies with $\log M_\star/M_\odot \ge 10$ which lie below 1$\sigma_\textrm{fit}$ of the mass-size relation, where $\sigma_\textrm{fit}$ is the standard deviation of the residuals from the fit. 

From these galaxies, we wish to select those with higher than average velocity dispersion. To this end, we measure $\sigma_e$, which we define as the the standard deviation from the best fitting Gaussian to the line-of-sight velocity distribution (LOSVD) of all particles inside one half-mass radius. The process of projecting the galaxies in two dimensions is explained later in subsection \ref{subsection:kin}. We proceed to fit the linear trend relating stellar mass and $\sigma_e$ of the galaxies selected in the previous step using a least squares approach and we selected the galaxies that lie above 1$\sigma_\textrm{fit}$ of this relation. Lastly, one object that was not tracked to high redshifts was then removed from the sample, which gives us a final sample size of 63 galaxies. The linear fits to the stellar mass-size and stellar mass-$\sigma_e$ relations are shown on the left panels of Fig. \ref{fig:sample_selec}.

To check if this sample selection procedure does select MCGs with similar properties to those studied by SM21, we applied it to the population of quiescent galaxies in the Sloan Digital Sky Survey (SDSS) DR15 \citep{aguado19}. We find that 27 galaxies selected in this manner were observed as part of the MaNGA survey, of which 25 were included in the study by SM21. While our selection method selects only 25 out of 70 galaxies studied by SM21, we note that these 25 galaxies follow the same trends in mass and stellar populations as the remaining objects. Thus, we argue that the selection criteria used in this paper is an adequate replacement to the criteria used by SM21.

\subsubsection{Control Sample Galaxies}

To build a control sample consisting of average-sized galaxies with average velocity dispersion for their stellar mass, we follow a procedure largely analogous to the compact galaxy sample selection.
We build our control sample by first selecting quiescent galaxies lying within $\pm 0.5 \sigma_\textrm{fit}$ of the mass-size relation. Next, we perform a linear fit to stellar mass-$\sigma_e$ distribution of the galaxies selected in the previous step and we select those within $\pm 0.5 \sigma_\textrm{fit}$. Again, this criteria was chosen as to be similar to the one adopted by SM21 to build their control sample of median sized galaxies.
The right panels of Fig. \ref{fig:sample_selec} show the two steps in the control sample galaxy (CSG) selection.
Finally, these galaxies were matched in $\sigma_e$ to the compact galaxies sample using the Propensity Score Matching method \citep{psm}. This ensures that both samples are of the same size and have a similar distribution in $\sigma_e$.

\subsection{Stellar Kinematics}
\label{subsection:kin}
\begin{figure*}
    \centering
    \includegraphics[width=\textwidth]{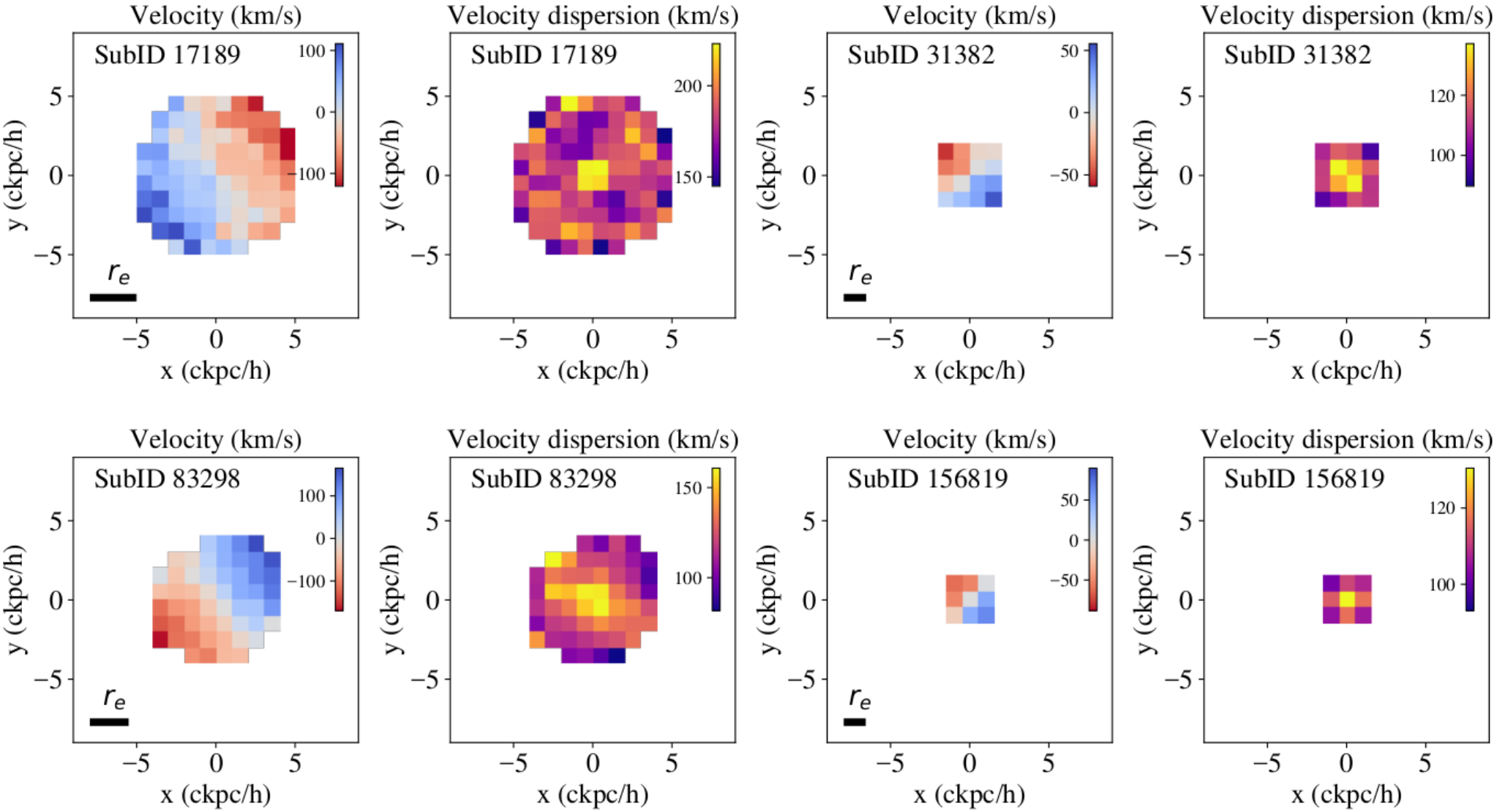}
    \caption{Velocity and velocity dispersion maps of four representative MCGs. The half-mass radius is drawn in the velocity maps as a solid black line. The galaxies depicted are: 
    $\mathrm{SubID}=17189$ ($\log M_\star = 11.0$, $R_e = 2.5$ kpc, $\sigma_e = 212.9$ km/s); 
    $\mathrm{SubID}=31382$ ($\log M_\star = 10.3$, $R_e = 0.9$ kpc, $\sigma_e = 139.4$ km/s); 
    $\mathrm{SubID}=83298$ ($\log M_\star = 10.6$, $R_e = 2.0$ kpc, $\sigma_e = 160.0$ km/s); 
            $\mathrm{SubID}=156819$ ($\log M_\star = 10.4$, $R_e = 0.9$ kpc, $\sigma_e = 142.2$ km/s).}
            \label{fig:maps}
            \includegraphics[width=\textwidth]{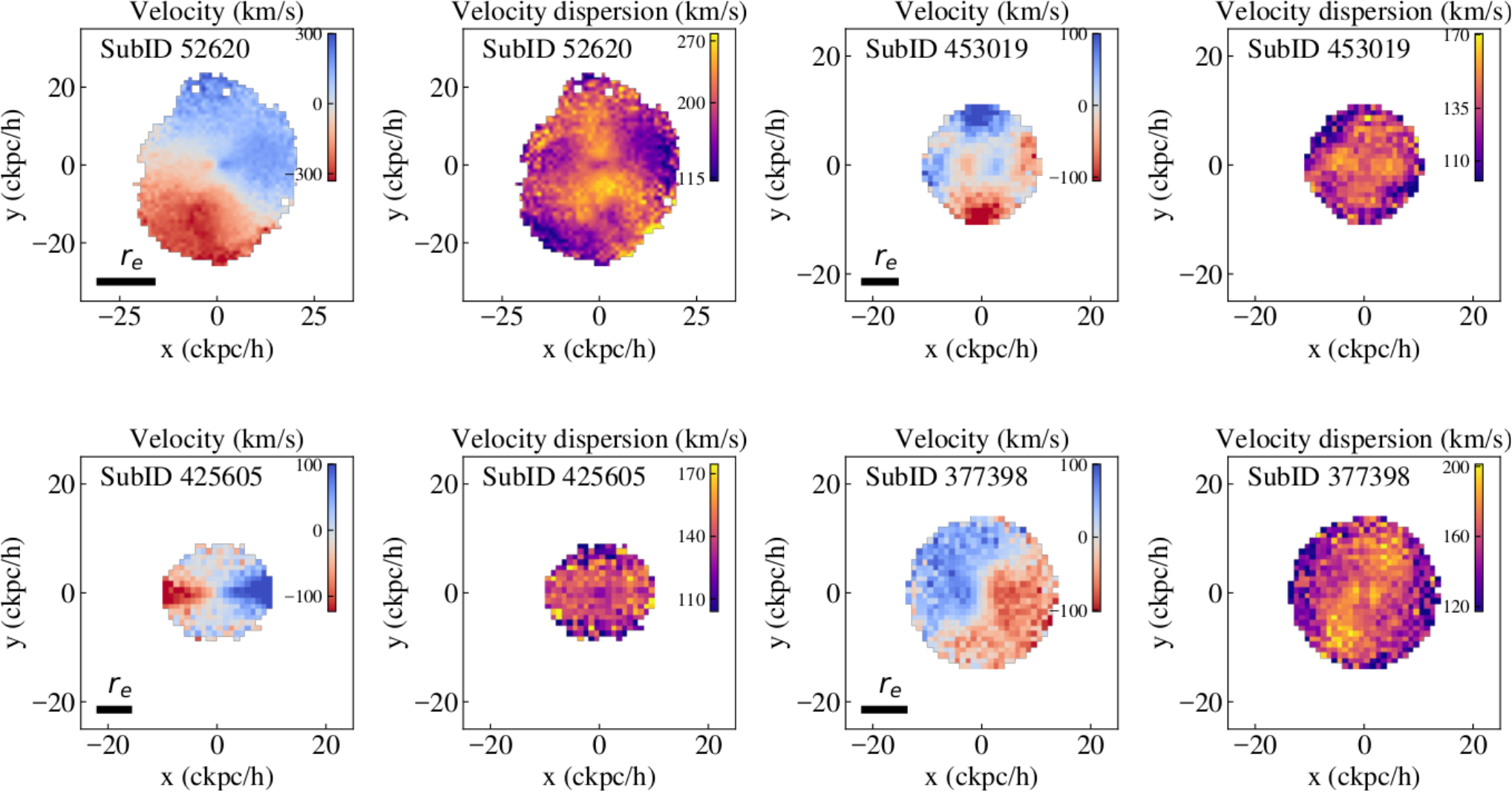}
            \captionof{figure}{Velocity and velocity dispersion maps of four representative CSGs. The half-mass radius is drawn in the velocity maps as a solid black line. The galaxies depicted are:
            $\mathrm{SubID}=52620$ ($\log M_\star = 11.5$, $R_e= 13.2$ kpc, $\sigma_e = 231.5$ km/s); 
            $\mathrm{SubID}=453019$ ($\log M_\star = 10.9$, $R_e= 5.5$ kpc, $\sigma_e = 147.0$ km/s); 
            $\mathrm{SubID}=425605$ ($\log M_\star = 10.8$, $R_e= 5.1$ kpc, $\sigma_e = 139.2$ km/s); 
            $\mathrm{SubID}=377398$ ($\log M_\star = 11.2$, $R_e= 7.1$ kpc, $\sigma_e = 173.8$ km/s).}
    \label{fig:maps2}
\end{figure*}

We measure the line-of-sight stellar velocity and velocity dispersion considering a random orientation of the galaxies by projecting the particles in the $xy$ plane and only considering their velocities in the $z$ direction, which we consider as the line of sight.  This corresponds to a random orientation since the simulation has no preferred direction. 

To construct stellar velocity and velocity dispersion maps for each galaxy, we divided the $xy$ plane into square bins of side length equal to twice the softening length of the simulation, $\epsilon = 0.74$ kpc. 
We then selected the particles inside each bin and looked at their LOSVDs, from which we extracted the velocity and velocity dispersion in the $i$-th bin as the weighted averages
\begin{equation}
    V_i=\frac{\sum_n M_{n,i}V_{n,i}}{\sum_n M_{n,i}}; \quad 
    \sigma_i=\sqrt{\frac{\sum_n M_{n,i}\left(V_{n,i}-V_i\right)^2}{\frac{N_i-1}{N_i}\sum_n M_{n,i}}},
\label{eq:Vsig}
\end{equation}
where the sum in $n$ runs over all the particles inside the $i$-th bin, $M_{n,i}$ is the $n$-th particle's mass, $V_{n,i}$ is their $z$ component velocity and $N_i$ is the number of particles contained in the $i$-th bin. In order to avoid spurious values in our maps, we imposed a minimum number of $N_\textrm{min}=10$ particles in each bin. Lastly, we remove the peculiar velocity of the galaxy in all bins by subtracting the median velocity of all particles of the galaxy. 

We quantify the angular momentum using the proxy parameter $\lambda_R$, defined as \citep{emsellem07}:
\begin{equation}
    \lambda_e=\frac{\sum_i M_i R_i|V_i|}{\sum_i M_i R_i\sqrt{V_i^2+\sigma_i^2}},
\end{equation}
where $R_i$ and $M_i$ are the circularized radius and the mass contained inside the $i$-th bin, respectively, and the sum is carried out over all bins in the kinematic maps. 
We choose to use masses instead of fluxes as weights for simplicity, and we verified that this choice has no big impact in the value of $\lambda_e$.

\subsection{Stellar populations}

We quantify the gradients following \citet{walters21}. We define an "inner" region of $r < r_{\rm e}$ and an "outer" region of $ r_{\rm e} < r < 2 r_{\rm e}$, and use the properties of the particles inside each region to calculate the gradients.
The mass weighted age gradient is given by
\begin{equation}
    \nabla_r \log \textrm{age} \equiv  \frac{\log \frac{\sum_{\textrm{out}} (m_\star) (\textrm{age}_\star)}{\sum_{\textrm{out}} (m_\star)} - \log \frac{\sum_{\textrm{in}} (m_\star) (\textrm{age}_\star)}{\sum_{\textrm{in}} (m_\star)} }{\log (1.5 r_{\rm e}) - \log (0.5 r_{\rm e})}\, ,
\label{gradage}
\end{equation}
whereas the metallicity gradient is
\begin{equation}
    \nabla_r \log [Z/H] \equiv  \frac{\log \frac{\sum_{\textrm{out}} m_{\star, Z}}{\sum_{\textrm{out}} m_{\star, H}} - \log \frac{\sum_{\textrm{in}} m_{\star, Z}}{\sum_{\textrm{in}} m_{\star, H}} }{\log (1.5 r_{\rm e}) - \log (0.5 r_{\rm e})}\, ,
\label{gradmetal}
\end{equation}
where $m_\star$ is the particle's stellar mass, $\textrm{age}_\star$ is the lookback time of its formation and $m_{\star, X}$ is the particle's mass fraction composed of element X. 
The subscripts "in" and "out" indicate the region in which the sum is performed. 
In addition to age and metallicity, we also calculated the gradients of $\alpha$ elements ($\nabla_r \log [\alpha/Fe]$) by restricting equation \ref{gradmetal} to only include $\alpha$ elements and replacing the hydrogen mass with that of iron.   
For this we used the abundances of the $\alpha$ elements that are traced in TNG, namely C, O, Ne, Mg and Si.

\subsubsection{Environment}

We characterise the global environment of a galaxy by the mass of its host halo. We say a galaxy belongs to a group if $M_\textrm{halo} < 10^{14} M_\odot$, and to a cluster if $M_\textrm{halo} \geq 10^{14} M_\odot$.
We characterise the local environment of galaxies using the tidal strength parameter $Q_\textrm{group}$, defined as:
\begin{equation}
\label{qgroup}
    Q_\textrm{group} \equiv \log \left[  \sum_{i=1}^{n-1} \frac{M_i}{M_P} \left( \frac{D_P}{d_i}\right)^3 \right]\, ,
\end{equation}
where $M_P$ and $D_P$ are the mass and the diameter of the primary galaxy, respectively, $M_i$ is the mass of the i-th neighbour and $d_i$ is its projected distance to the primary galaxy. This parameter quantifies the strength of tidal forces produced by neighbor galaxies, such that a galaxy is considered isolated if these forces are smaller than 1\% of its internal binding force, i.e. $Q_\textrm{group}<-2$ \citep{athanassoula84, verley07}.

\section{Results}

\subsection{Stellar Kinematics}

In Fig. \ref{fig:maps} and Fig. \ref{fig:maps2} we show the velocity and velocity dispersion maps of four MCGs and four CSGs respectively, where each galaxy is representative of objects with similar stellar mass in their respective sample. 

In the case of compact galaxies, we notice that due to their small sizes (adopted as the half-mass radius), which in some cases is comparable to the simulation's softening length $\epsilon=0.74\,$kpc, the kinematic maps have few bins. Still, a rotation pattern with high velocity dispersion in the central region is observed in the majority of the objects, suggesting simulated MCGs are bulge+disk systems, in qualitative agreement with observations.
Compared to MCGs, CSGs are significantly larger and therefore better resolved, allowing for a more detailed analysis of their kinematic structures. The stellar velocity maps are usually complex, but a rotation pattern is frequently present, although iso-velocity contours are commonly twisted. Velocity dispersion maps show no clear pattern, and unlike what is seen in MCGs, the highest velocity dispersion values are not restricted to the center of the galaxy.

\begin{figure}
    \centering
    \includegraphics[width=\columnwidth]{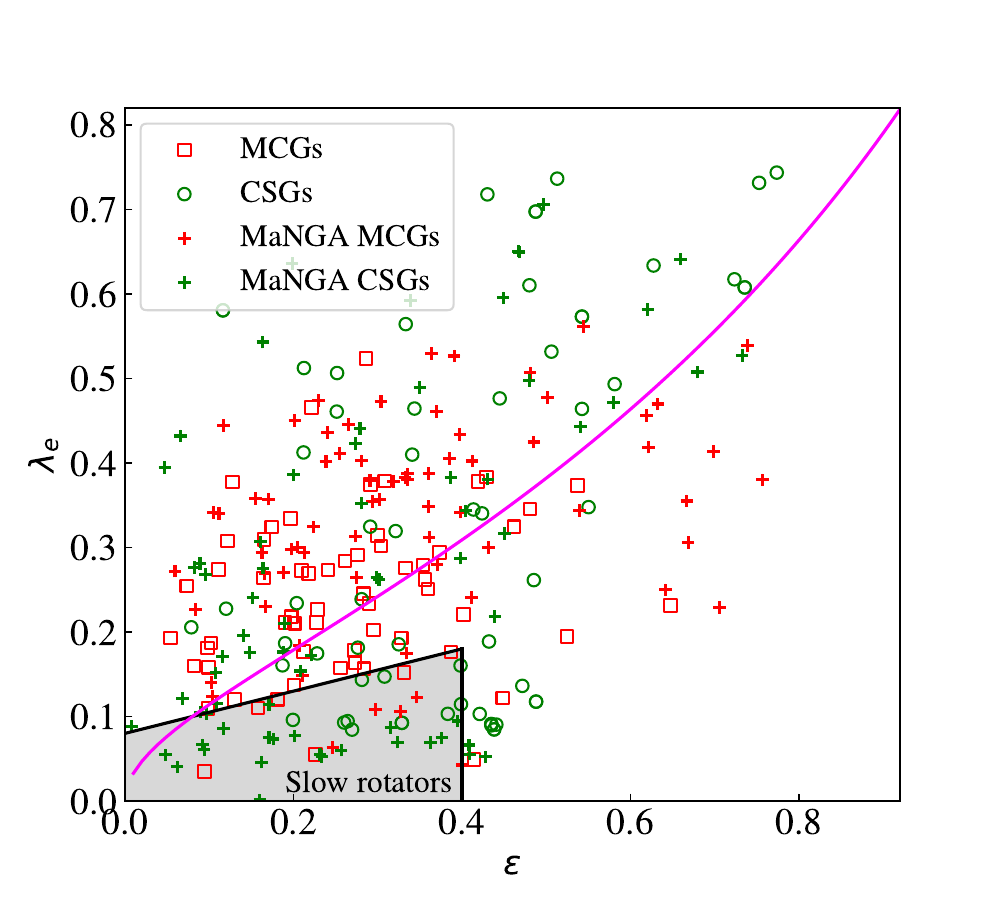}
    \caption{$\lambda_e$ versus ellipticity for the TNG compact sample (open red squares) and control sample (open green circles). The magenta line represents the relation for edge-on galaxies with a constant anisotropy factor $\delta = 0.7 \epsilon$. The shaded area indicates the region in the diagram occupied by slow rotators, as defined in equation \ref{eq:slow}. For comparison we show MaNGA MCGs and CSGs as red and green crosses respectively.}
    \label{fig:lambda_eps} 
\end{figure}

Fig. \ref{fig:lambda_eps} shows the distribution of the two samples in the $\lambda_R-\epsilon$ diagram. The magenta line represents the expected values of $\lambda_R$ for a galaxy of ellipticity $\epsilon$ and a constant anisotropy factor $\delta = 0.7 \epsilon$ (see  \citealt{cappellari07} for details).
Galaxies are classified as slow or fast rotators, depending on the value of the spin parameter $\lambda_e$ for a given ellipticity. Slow rotators are defined as having \citep{cappellari16}:

\begin{equation}
\label{eq:slow}
     \lambda_e < 0.08 + \epsilon/4 \quad \textrm{with} \quad \epsilon < 0.4\, ,
\end{equation}
Slow rotators occupy the shaded region in Fig. \ref{fig:lambda_eps}. 

Both MCGs (90.5\%) and CSGs (82.5\%) are predominantly fast rotators, but they occupy different regions of the $\lambda_e-\epsilon$ diagram. MCGs tend to have lower ellipticities ($\epsilon \lesssim 0.6$) and spins ($\lambda_e \lesssim 0.5$), typical of bulge dominated objects. CSGs instead are a diverse group in terms of kinematics, composed of quiescent disk/bulge dominated galaxies as well as slowly rotating ellipticals, as evidenced by the wide range of ellipticities and spins. A comparison with MaNGA MCGs and CSGs (red and green crosses in Fig.\,\ref{fig:lambda_eps}) show that although the fast rotator fraction of simulated MCGs is in reasonable agreement with observations, simulated MCGs tend to have lower $\lambda_e$, and there is a clear lack of simulated MCGs in the $\lambda_e \sim 0.4-0.6$ range compared to observations.

\subsection{Environment}

\begin{figure}
    \centering
    \includegraphics[scale=0.50]{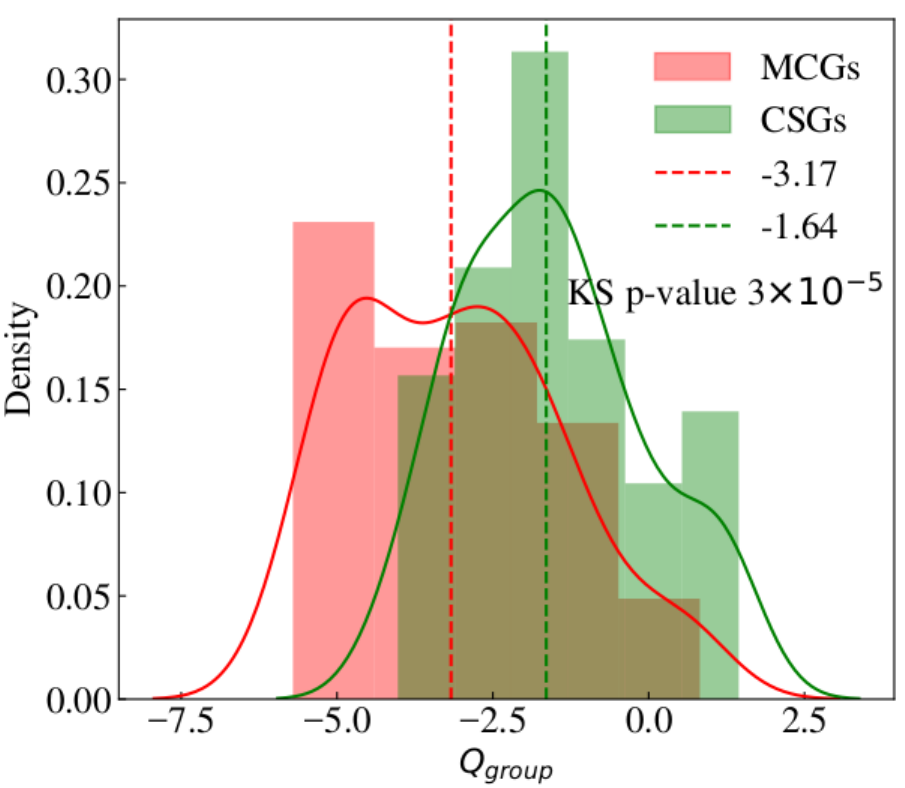}
    \caption{Distributions of the local environment parameter $Q_\mathrm{group}$ for compact galaxies (red) and control galaxies (green). Solid lines represent kernel density estimates using Gaussian kernels with adaptive bandwidths following Scott's rule \citep{scottrule}, while dashed lines show the medians of the distributions.}
    \label{fig:qgroup}
\end{figure}

\begin{figure}
     \centering
     \includegraphics[scale=0.50]{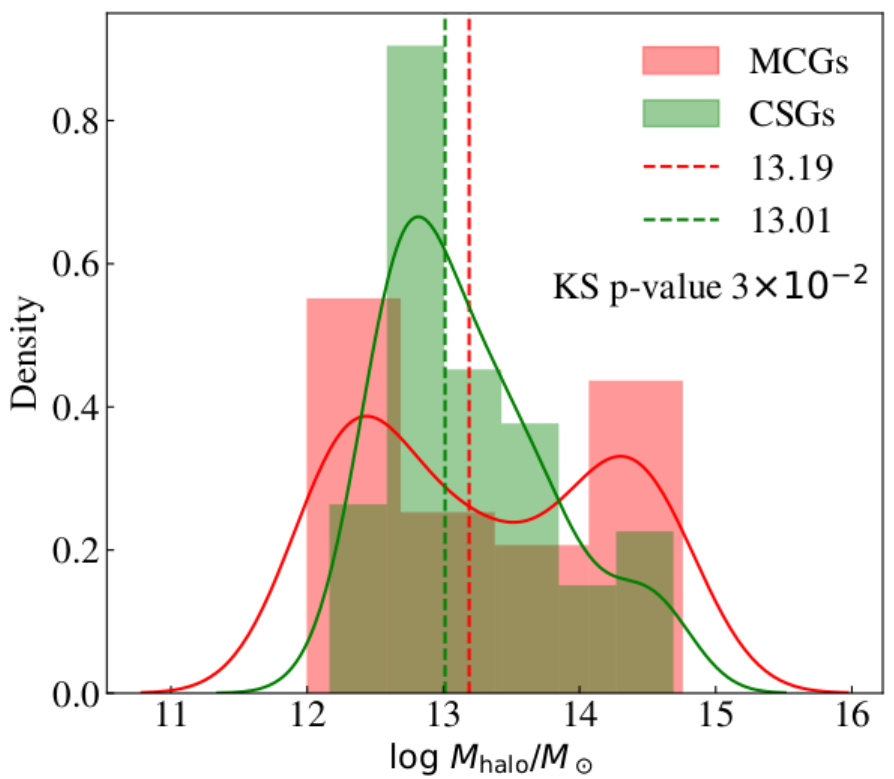}
     \caption{Mass distributions of the dark matter halos that harbor our compact galaxies (red) and control galaxies (green). Solid lines represent kernel density estimates using Gaussian kernels with adaptive bandwidths following Scott's rule \citep{scottrule}, while dashed lines show the medians of the distributions.}
     \label{fig:halo_mass}
\end{figure}

\renewcommand{\arraystretch}{1.5}
\begin{table*}
\centering
\caption{The number of galaxies in halos within different mass bins are divided in central and satellites. The last line show the $p$-values of Fisher's test comparing the fractions of compact and control sample galaxies that are centrals and satellites in each $M_\mathrm{halo}$ bin.}
\begin{tabular}{c|cc|cc|cc|c}
\hline
\multicolumn{1}{c|}{} & \multicolumn{2}{c|}{$M_{\rm halo}$\,$<$\,10$^{13}$\,M$_{\odot}$} & 
\multicolumn{2}{c|}{10$^{13}$\,$\leq$\,$M_{\rm halo}$\,$<$\,10$^{14}$\,M$_{\odot}$} & 
\multicolumn{2}{c|}{$M_{\rm halo}$\,$\geq$\,10$^{14}$\,M$_{\odot}$} & Total\\
\cline{2-8}
& Centrals  & Satellites & Centrals & Satellites & Centrals & Satellites & \\ 
\hline \hline
\multirow{2}{*}{MCGs} & 27 (96.4\%) & 1 (3.6\%) & {5 (38.5\%)} & {8 (61.5\%)} & {0 (0\%)} & {22 (100\%)} & \\
    & \multicolumn{2}{c|}{\cellcolor{blue!20} {28 (44.4\%)}} &                          \multicolumn{2}{c|}{\cellcolor{blue!20} {13 (20.6\%)}} & 
    \multicolumn{2}{c|}{\cellcolor{blue!20} 22 (34.9\%)} & 
    \multicolumn{1}{c}{\cellcolor{blue!20} 63} \\  
    \hline       
\multirow{2}{*}{CSGs} & 30 (96.8\%) & 1 (3.2\%) & {14 (58.3\%)} & {10 (41.7\%)} & { 1 (12.5\%)} & {7 (87.5\%)} & \\ 
    & \multicolumn{2}{c|}{\cellcolor{blue!20} {31 (49.2\%)}} &             \multicolumn{2}{c|}{\cellcolor{blue!20} {24 (38.1\%)}} &     
    \multicolumn{2}{c|}{\cellcolor{blue!20} 8 (12.7\%)} & 
    \multicolumn{1}{c}{\cellcolor{blue!20} 63} \\
    \hline         

    \multicolumn{8}{c}{\bf \large Fisher's tests} \\
    \hline \hline
    
    \multirow{2}{*}{$p$-value} & 0.7 & 0.7 & 0.2 & 0.2 & 0.3 & 0.3 & \\

    & \multicolumn{2}{c|}{\cellcolor{blue!20} {0.7}} &                          \multicolumn{2}{c|}{\cellcolor{blue!20} {$5 \times 10^{-2}$}} &     
    \multicolumn{2}{c|}{\cellcolor{blue!20}  $6 \times 10^{-3}$} & 
    \multicolumn{1}{c}{\cellcolor{blue!20} } \\
    
    
    \hline
\end{tabular}
\label{table:env}
\end{table*}
\renewcommand{\arraystretch}{1}

In Fig. \ref{fig:qgroup} we show the distribution of the tidal strength parameter $Q_\textrm{group}$.
We define the condition of $Q_\textrm{group}<-2$ for a galaxy to be considered isolated, and we can see that the majority (74.6\%) of compact galaxies are below this value. The control sample, however, has a significant fraction of galaxies above this cutoff (57.1\%). This indicates that compact galaxies in the simulation tend to be in a more locally isolated environments than median-sized galaxies.

Fig. \ref{fig:halo_mass} shows the distributions of host dark matter halo mass. Although the distributions have similar medians, their shapes are unlike. MCGs appear to have a double peaked distribution, being more common in low mass ($\log M_\mathrm{halo} \sim 12.5$) and cluster-sized halos ($\log M_\mathrm{halo} \geq 14.0$). In contrast, CSGs tend to be hosted by halos with $\log M_\mathrm{halo} \sim 13.0$, and the distribution is asymmetric with a tail extending to larger masses. 

In Table \ref{table:env} we show the fractions of central/satellite galaxies in each sample for three halo mass bins. There are no statistically significant differences between the samples: both CSGs and MCGs tend to be centrals when hosted by low mass halos, those inhabiting group sized halos they are a mix of central and satellites, and those in cluster sized halos are satellites.

\subsection{Stellar Populations}

\begin{figure*}
    \centering
    \includegraphics[scale=0.55]{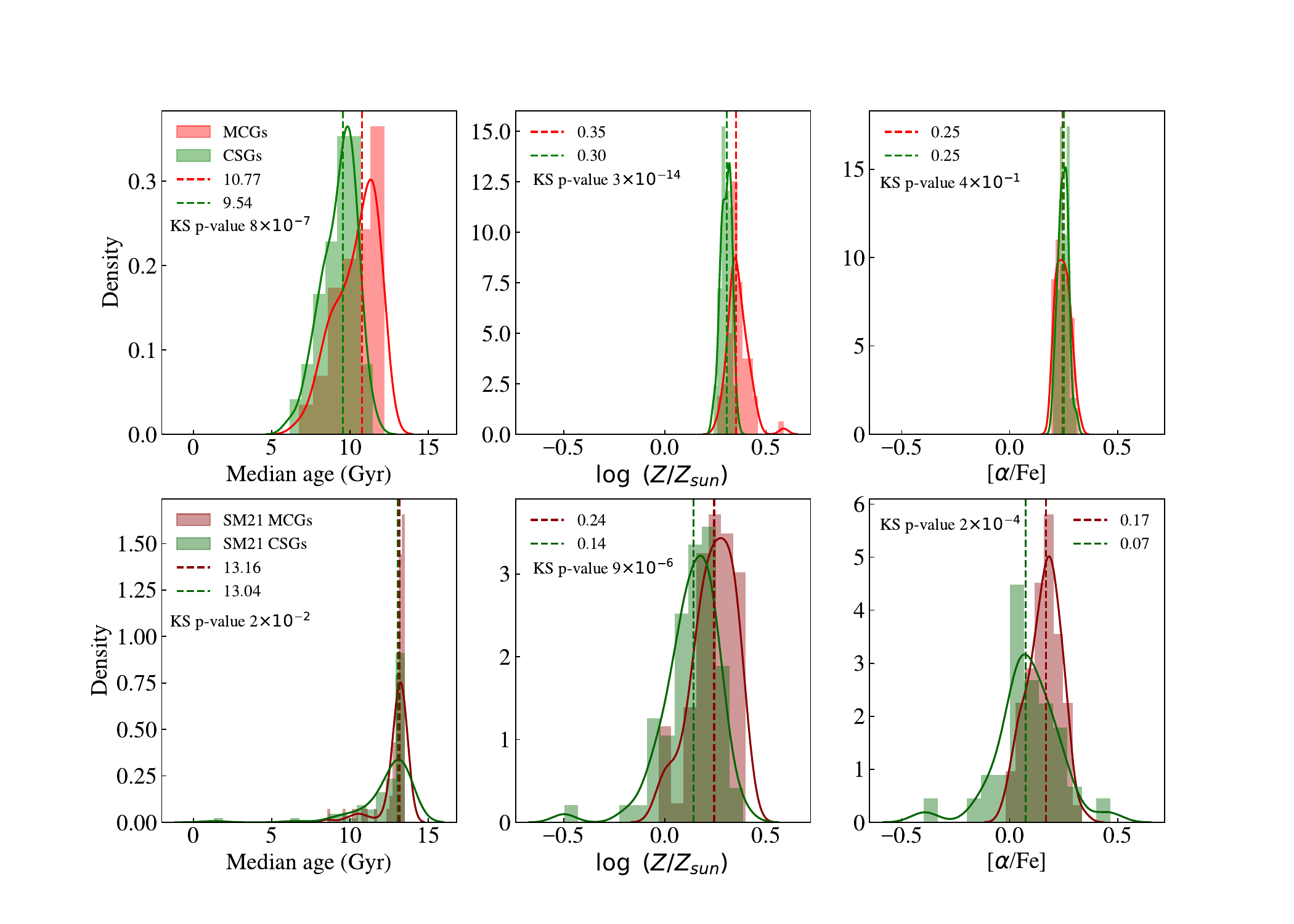}
    \caption{Distributions of several quantities characterizing the stellar populations of MCGs (red) and CSGs (green). The top panels show the distributions of median age, metallicity and [$\alpha$/Fe] calculated inside one half-mass radius for our TNG samples, whereas the bottom panels show the same quantities for observed MaNGA MCGs and CSGs from SM21. In all panels, the solid lines represent kernel density estimates using Gaussian kernels with adaptive bandwidths following Scott's rule \citep{scottrule}, while the dashed lines indicate the medians of the distributions.}
    \label{fig:age_metal_alpha_inner}
\end{figure*}

\begin{figure*}
    \centering
    \includegraphics[scale=0.55]{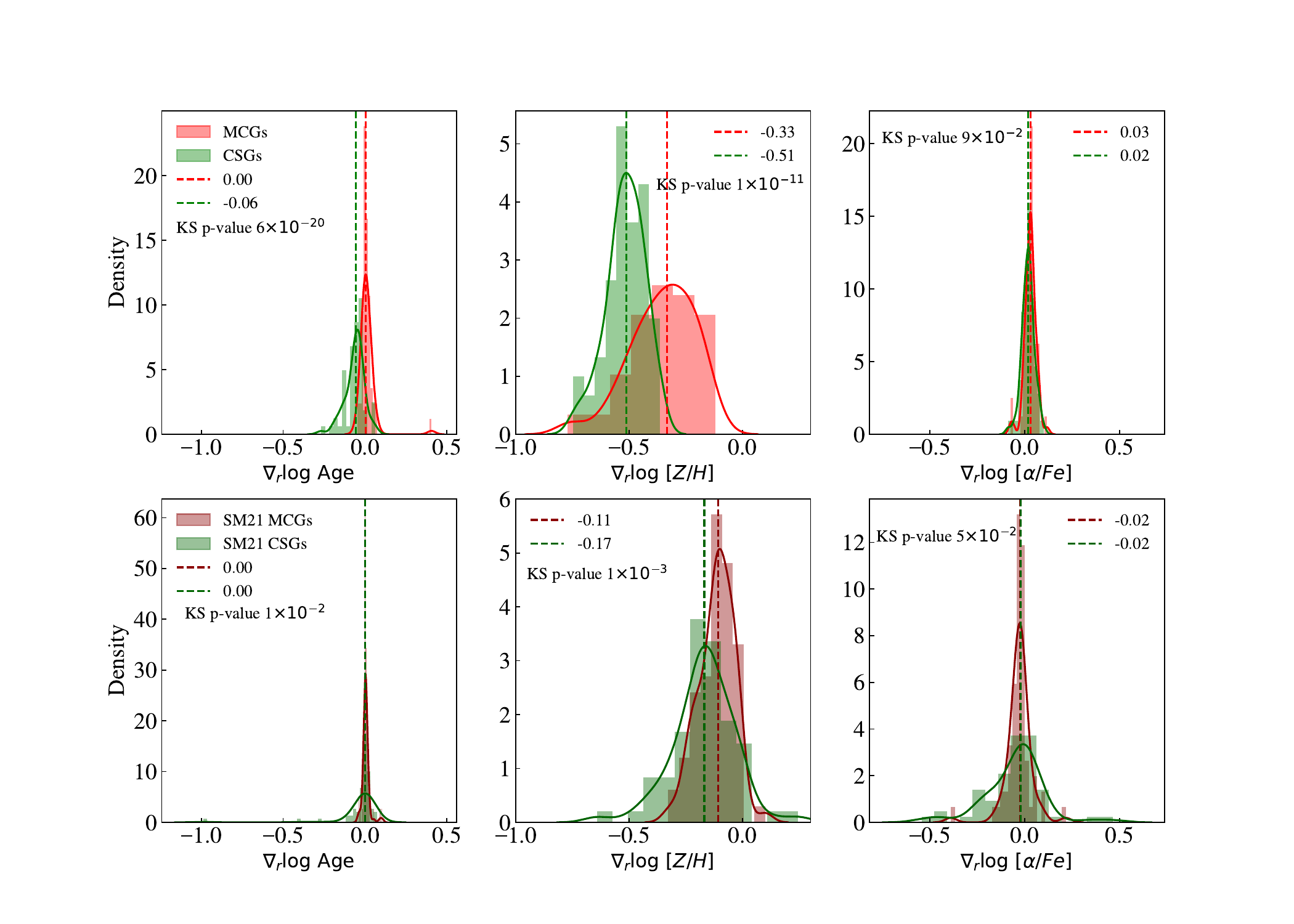}
    \caption{Distributions of the stellar populations gradients of MCGs (red) and CSGs (green). The top panels show the gradients of age, metallicity and [$\alpha$/Fe] for our TNG samples, while the bottom panels show the gradients of the SM21 MaNGA MCGs and CSGs. The gradients were obtained considering an inner region of $R < R_e$ and an outer region of $R_e < R < 2 R_e$ following equations \ref{gradage} and \ref{gradmetal}. In all panels, the solid lines represent kernel density estimates using Gaussian kernels with adaptive bandwidths following Scott's rule \citep{scottrule}, while the dashed lines indicate the medians of the distributions.}
    \label{fig:age_metal_alpha_grad}
\end{figure*}

The top panels in Fig. \ref{fig:age_metal_alpha_inner} show the age, metallicity and $\alpha$-enhancement of MCGs and CSGs. In terms of age, both distribution have a similar shape, with a peak at old ages and a long tail extending to younger ages. MCGs are older, with a median age of $\sim 10.8$ Gyr compared to $\sim 9.5$ Gyr for CSGs. The age distribution of MCGs is also broader. MCGs and CSGs have supersolar metallicities, with MCGs being more metal rich. MCGs and CSGs are $\alpha$-enhanced, both having similar median [$\alpha$/Fe] values and distribution widths. In the bottom panels we show, for comparison, the stellar population properties of the MaNGA MCGs and CSGs studied by SM21. 
The observed and simulated mass-weighted metallicities distributions are in reasonable qualitative agreement. 
The observed mass-weighted ages are significantly larger than the simulated, however this is not surprising considering that it is very difficult to measure ages for stellar populations older than $\gtrsim 9$\,Gyr \citep{conroy13}, so they are likely overestimated. The simulation reproduces the observed trends of MCGs being older and having higher metallicities than CSGs, although it fails in reproducing the observed differences in [$\alpha$/Fe].

The top panels of Fig. \ref{fig:age_metal_alpha_grad} show the distributions of age, metallicity and $\alpha$-enhancement gradients for the two samples. MCGs have a narrow age gradient distribution with a median consistent with zero. CSGs, on the other hand, show a significantly wider distribution.  Metallicity gradients are negative for both samples, but the MCGs distribution is broader and the median is larger. Lastly, the gradients of $\alpha$-enhancement elements of both samples are similar. In the bottom panel we show, for comparison, the stellar population gradients for MaNGA MCGs and CSGs. Simulated and observed age and [$\alpha$/Fe] gradients are similar, on the hand simulated metallicity gradients are significantly steeper for both MCGs and CSGs.

As some MCGs have sizes of the order of the softening length of the simulation, there is a possibility that the differences in the gradientes of MCGs and CSGs are mainly caused by differences in resolution. To test this, we first check if the stellar population gradients of MCGs show a significant correlation with size. We find that $[\alpha/Fe]$ gradients show a moderate correlation with size, while the metallicity gradients are strongly correlated. As a second test, we select samples of MCGs and CSGs in TNG50, which has a much higher resolution of $\epsilon = 0.288$. The samples are very small, only eight objects in each, nonetheless we recover similar trends for the stellar populations inside 1 $R_e$ and for the age and [$\alpha$/Fe] gradients, but not for the metallicity gradient, which are steeper in TNG50, with MCGs and CSGs having similar values. The stellar population gradients of the TNG50 sample are shown in Appendix\,\ref{appendix_a}. Thus, it is not possible to assess the reliability of the metallicity gradients, we refrain from discussing them from here on. 

\subsection{Star Formation History}

\begin{figure*}
    \centering
    \includegraphics[scale=0.55]{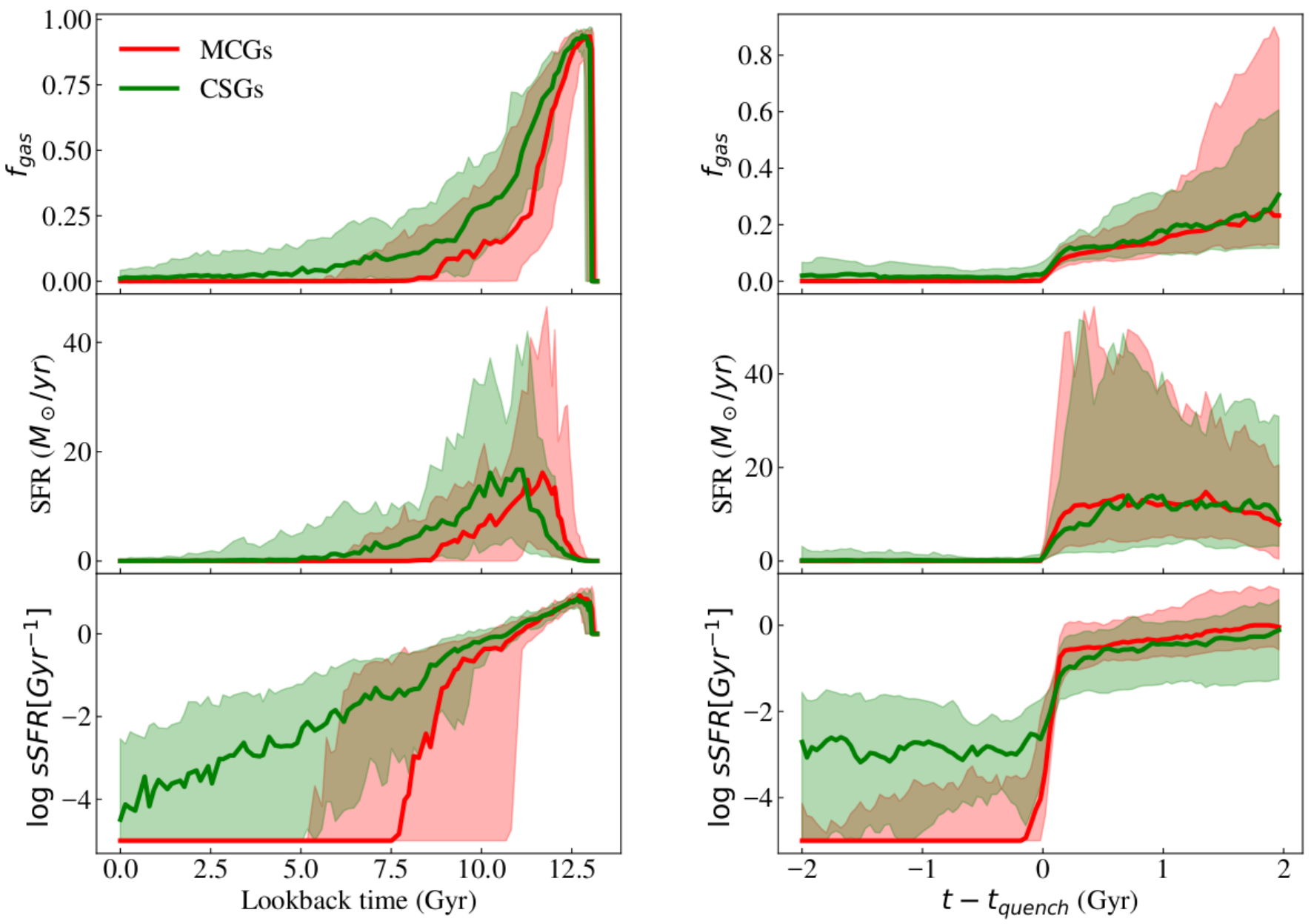}
    \caption{Time evolution of the gas fraction $f_\mathrm{gas}$ (top), the star formation rate (middle) and the specific star formation rate (bottom) for the MCGs (red) and CSGs (green)}. On the left, the evolution is shown as a function of lookback time, while on the right it is plotted as a function of the time before each galaxy quench. On the lower panels, galaxies with zero star formation are plotted as having $\log\, sSFR = -5$ for visualization purposes. The shaded areas correspond to the 16th and 84th percentiles of the distributions.
    \label{fig:bhdot_mgas_sfr}
\end{figure*}

To determine the redshift where a given galaxy quenched star formation, $z_\mathrm{quench}$, we considered the mean specific star formation rate of galaxies with mass $10^9 < M_\star/M_\odot < 10^{10.5}$ and $SFR>0$, the so called ridge of the star-formation main sequence, in every snapshot. We consider as the quenching snapshot the one where the specific star formation rate falls below 1 dex of the ridge of the main sequence, and remains so for at least 5 snapshots.
The gas fraction, SFR and specific star formation rate (sSFR) as a function of lookback time (left column) and time before quenching (right column) are shown in Fig. \ref{fig:bhdot_mgas_sfr}.  In terms of lookback time, the SFR, the gas fraction and the sSFR starts to decline earlier in MCGs, indicating that, on average, MCGs form earlier than CSGs. In particular, the SFR of MCGs rise very fast in the beginning of the simulation, peaking by $z \sim 3-4$, while CSGs experience a significantly slower rise in SFR, reaching a peak by $z \sim 2$.

When plotted as a function of time before quenching, there are small differences in the median SFR curves. MCGs experience a sharp decline in SFR starting a few hundred million years before $t_\mathrm{quench}$, while the decline in the SFR of CSGs is gradual and starts earlier. The median sSFR curves show a similar trend to the SFR curves, although the decline in the sSFR of MCGs is sharper than the decline in SFR. In terms of the gas fraction, there are no significant differences in the rate in which the gas fraction decreases in MCGs and CSGs.

\subsection{Supermassive Black Hole Accretion History}

\begin{figure*}
    \centering
    \includegraphics[scale=0.55]{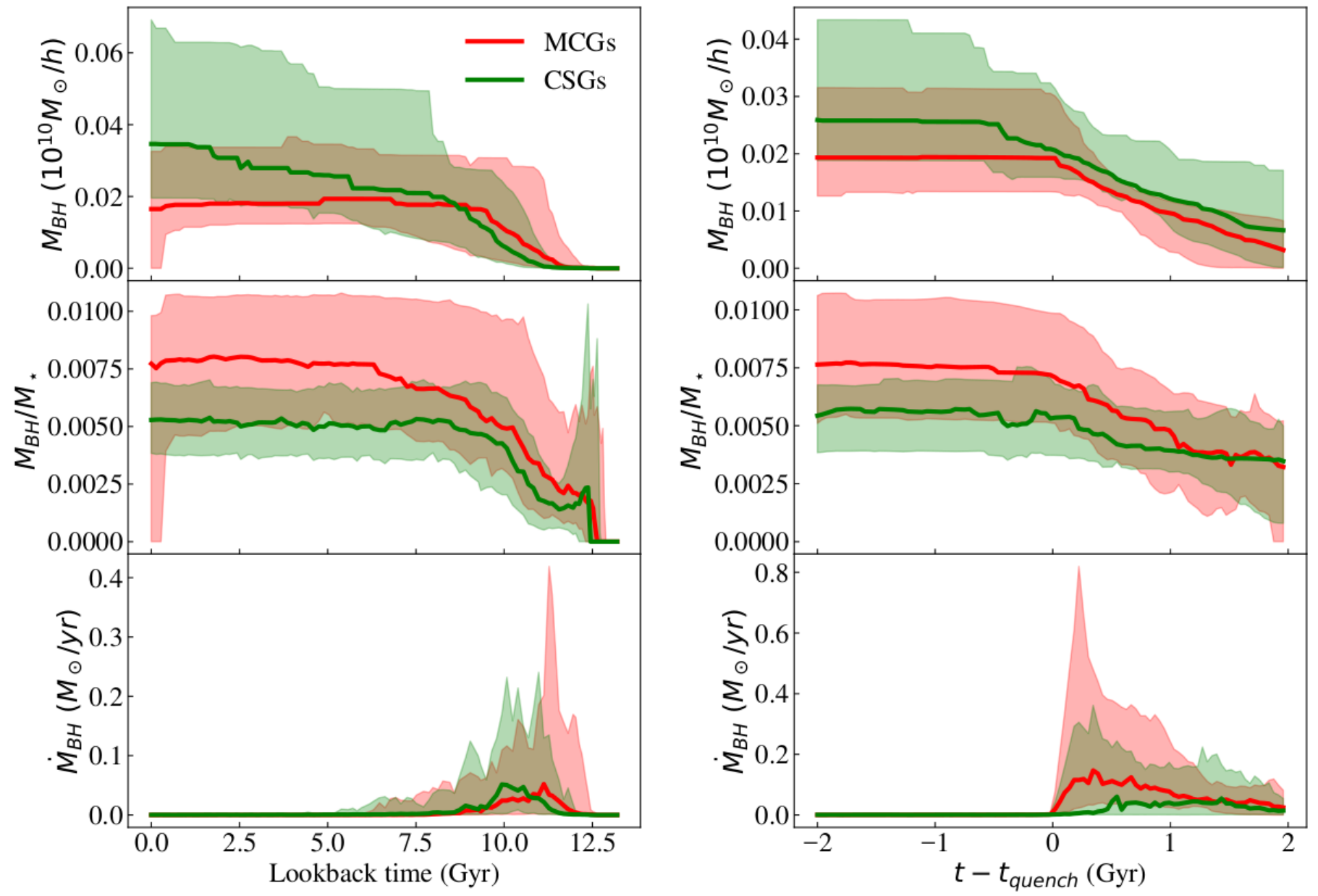}
    \caption{Analogous to figure \ref{fig:bhdot_mgas_sfr}. but now showing SMBH properties. From top to bottom: SMBH mass, SMBH mass normalised by stellar mass and the SMBH accretion rate.}
    \label{fig:bh_mass}
\end{figure*}

\begin{figure}
    \centering
    \includegraphics[scale=0.45]{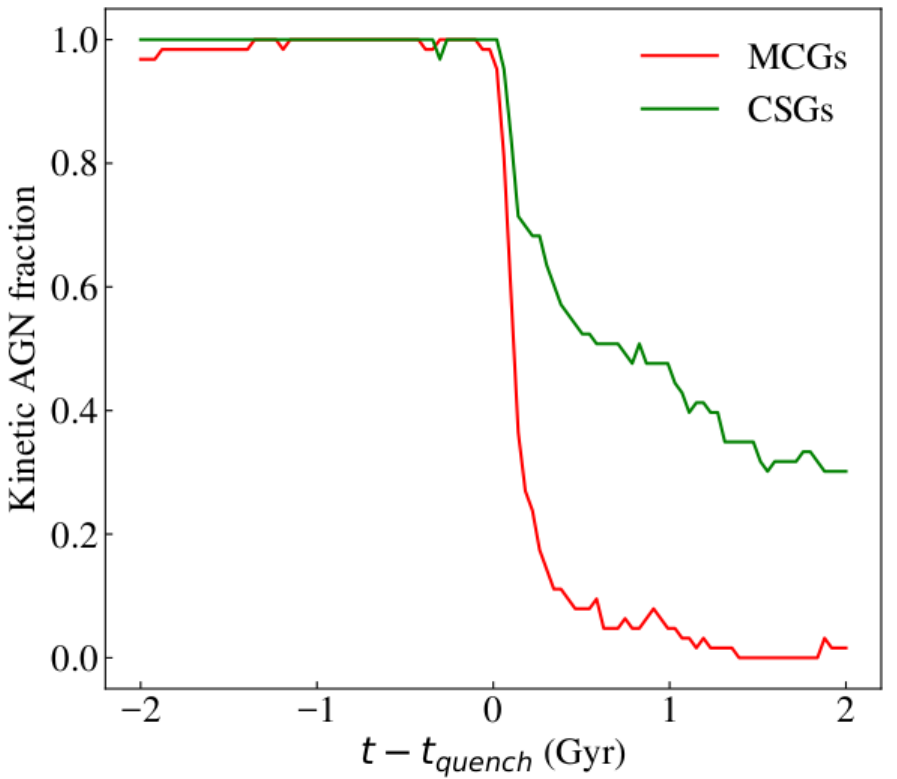}
    \caption{The fraction of galaxies with the central AGN in kinetic mode as a function of time before each galaxy quench, with the red line showing MCGs and the green line showing CSGs. The AGN mode was obtained using the definition discussed in \citet{weinberger17}.}
    \label{fig:kinetic_agn}
\end{figure}

It is known that, for massive galaxies in IllustrisTNG, the main mechanism for quenching star formation is AGN feedback in kinetic mode \citep{weinberger18, donnari21}, when the supermassive black hole (SMBH) is in a low accretion state. Kinetic mode feedback acts as both ejective feedback and preventative feedback, as it expels gas from the galaxy and heats the circumgalactic medium. Despite supplying most of the feedback energy, the thermal mode appears to be relatively ineffective \citep{zinger20}. In IllustrisTNG the dividing line between quasar-mode and kinetic-mode feedback depends on both the Eddington ratio and the SMBH, such that SMBHs with masses $\gtrsim 10^{8.5}\,M_\odot$ are mostly in kinetic mode \citep{terrazas20}. 

Fig. \ref{fig:bh_mass} shows the evolution of the SMBH mass for both samples. The top panels show the SMBH mass, the middle panels shows the ratio of SMBH mass to stellar mass and the bottom panels show the accretion rate.  In the left panels, these properties are shown as a function of lookback time, while in the right panels they are shown as a function of time before quenching. In the left column, we see that the SMBH in MCGs start to grow before those in CSGs, being more massive until a lookback time of $ \sim 9$ Gyr when the SMBH in CSGs become more massive. Note, however, that the ratio of SMBH mass to stellar mass is larger in MCGs throughout the entire evolution of the simulation. In the right column of Fig. \ref{fig:bh_mass}, an interesting behavior can be seen in evolution of $\dot{M}_\mathrm{BH}$ and $M_\mathrm{BH}/M_\star$: starting around 1\,Gyr prior to $t_\mathrm{quench}$, both show an increase in MCGs, pointing to a period of increased AGN activity before quenching in these galaxies. In contrast, while $M_\mathrm{BH}/M_\star$  mildly increases in CSGs over the same period, $\dot{M}_\mathrm{BH}$ declines. 

In Fig. \ref{fig:kinetic_agn} we show the fraction of galaxies with the SMBH in kinetic mode as a function of the time of quenching. For MCGs, the fraction steeply increases from $\sim 5-10\%$ at 1\,Gyr before quenching to 100\% shortly before quenching, coinciding with a steep drop in the sSFR. At 1\,Gyr before quenching $\sim 30-40\%$ of CSGs are in kinetic mode feedback, the fraction increasing gradually. This can explain the slower decline in the median SFR and sSFR of CSGs.

\subsection{Merger history}

\begin{figure*}
    \centering
    \includegraphics[scale=0.47]{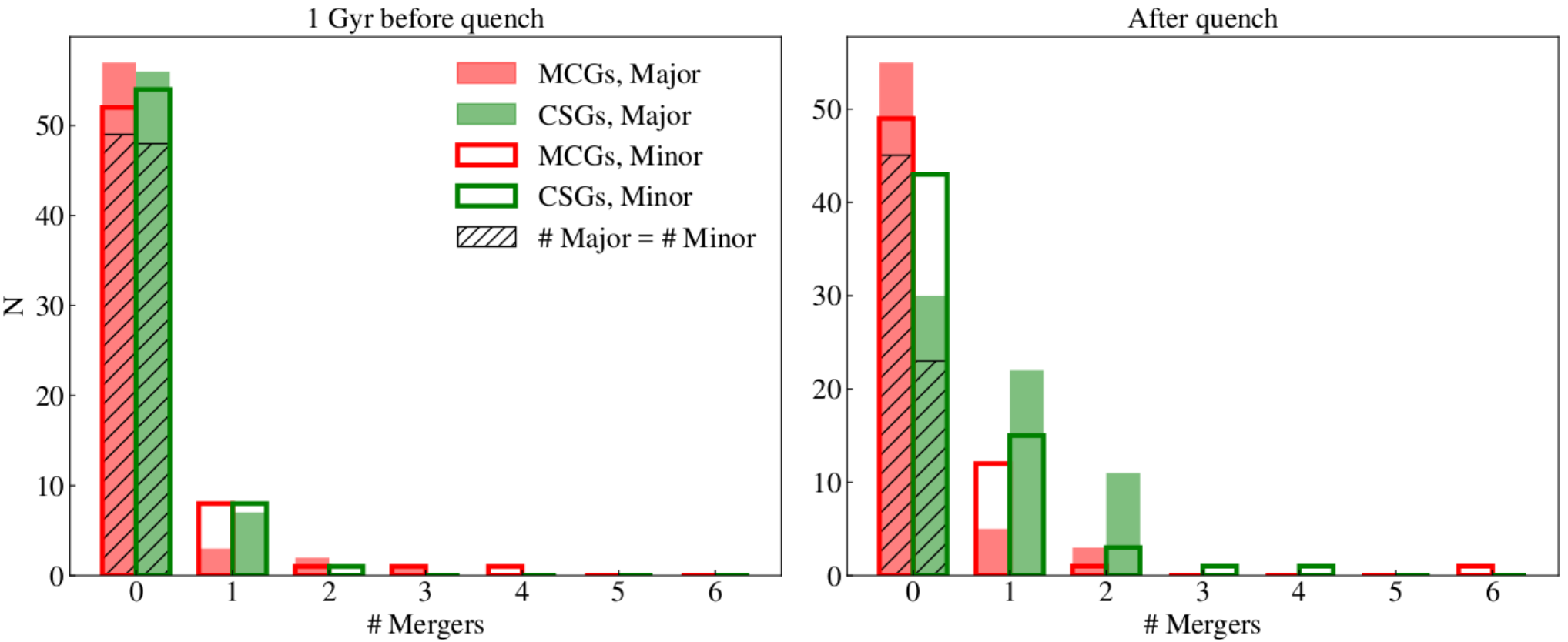}
    \caption{Number of mergers undergone by MCGs (red bars) and CSGs (green bar). On the left panel, we show the number of mergers that happened 1 Gyr before each galaxy quench, while on the right panel we show the number of mergers that happened after quenching. Filled bars depict major mergers, unfilled bars show minor mergers, and the hatched bars show the number of galaxies that did not suffer mergers of any kind.}
    \label{fig:majMergers}
\end{figure*}

\begin{figure}
    \centering
    \includegraphics[scale=0.5]{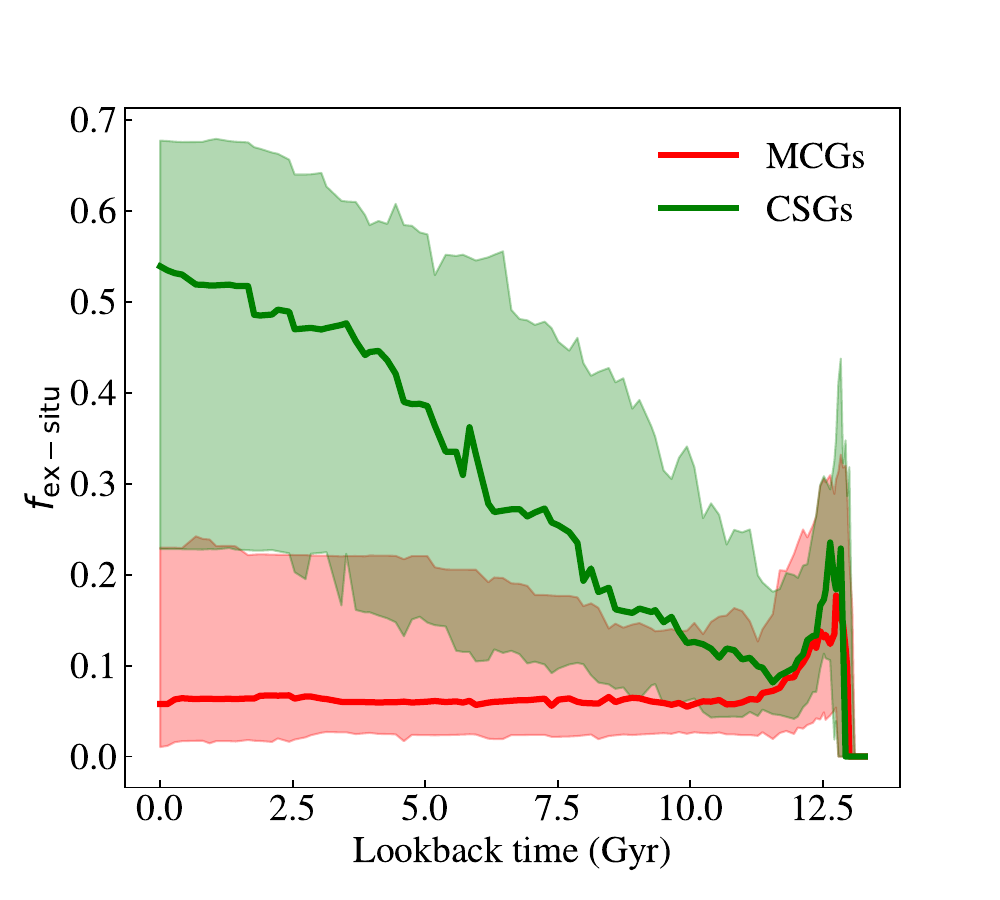}
    \caption{Evolution of the ex situ mass fraction $f_\mathrm{ex-situ}$ obtained from the stellar assembly supplementary catalog \citep{rodriguez-gomez16} for MCGs (red) and CSGs (green).}
    \label{fig:ex-situ}
\end{figure}

It has been proposed that major mergers are capable of triggering the quenching of star formation in galaxies, as tidal torques drive large amounts of gas to the center of the galaxy, enhancing star formation and AGN activity \citep{dimatteo05,hopkins08,pontzen17,davies22}. Observations of post-starbursts galaxies support a connection between galaxy mergers and fast star formation quenching, as they frequently display tidal features and disturbed morphologies \citep{sazonova21,ellison22,verrico22}. Furthermore, a large fraction of $z \sim 2$ fast quenching massive compact galaxies were found to reside in overdensities \citep{belli19}, which suggests mergers play an important role in their formation. 

To explore the role of mergers in the quenching of star formation in MCGs, we show in the left panel of Fig. \ref{fig:majMergers} the number of mergers that MCGs and CSGs underwent in the last one billion year before becoming quiescent. We can see that the behavior of the two samples is similar: most MCGs and CSGs did not undergo any merger, suggesting that mergers are not the main triggers of quenching. This is in line with previous studies that reported quenching is rare in post-merger galaxies in IllustrisTNG \citep{quai21}. 

The main reason why MCGs at low and intermediate redshift have been the subject of study is that they are local analogues of high redshift compact quiescent galaxies. However, their usefulness as analogues depends on the amount of mass they have accreted since quenching. An analysis of the stellar kinematics of MCGs suggested that the majority have had a quiet merger history (SM21), but a firm confirmation that a galaxy accreted little to no mass since formation requires either an analysis of the stellar population properties up to several $R_e$ - which require very long exposures to reach a high signal-to-noise ratio in the faint outskirts of the galaxy - or an analysis of the optical colors of the globular cluster system - which requires very high spatial resolution and thus is only possible for nearby galaxies. Because of this, as of this moment, only a few nearby objects have been confirmed to have accreted negligible amounts of mass since quenching (see \citealt{yildirim17} and \citealt{beasley18} for examples). In this context, cosmological simulations such as IllustrisTNG offers an advantageous alternative to assess the post-quenching merger history of MCGs.

In the right panel of Fig. \ref{fig:majMergers} the number of mergers a galaxy experienced after becoming quiescent is shown. A clear difference between the samples is apparent, with 28.6\% of MCGs experiencing mergers compared to 63.5\% of CSGs. Of the MCGs that experience a merger after quenching, 44.4\% experience a major merger and 77.8\% experience a minor merger. For CSGs the percentages are 82.5\% and 50\% respectively. We find that the fraction of MCGs that experience mergers show a small variation with age: 38.1\% (8 out of 21) of MCGs that quenched earlier than $z = 2$, 30.4\% (7 out of 23) for those that quenched between $1 \le z < 2$ and 15.8\% of those that quenched at $z < 1$ experienced a merger. In terms of host halo mass, we find little variation in the fraction of galaxies that have merged. The percentage of MCGs that have merged are 31.8\% in massive halos ($\log M_\mathrm{halo} \ge 14$), 30.8\% in intermediate mass halos ($13 \le \log M_\mathrm{halo} < 14$) and 25\% in low mass halos ($\log M_\mathrm{halo} < 13$).

In Fig. \ref{fig:ex-situ} we show the fraction of stars formed ex-situ as a function of lookback time. After a period of increased merging activity in the beginning of the simulation, the median ex-situ fraction of both MCGs and CSGs goes through a period of decline which lasts until $\sim 11$ Gyr ago. After, they follow distinct trajectories. In MCGs the median ex-situ fraction stabilizes at $\sim 7\%$, while it continuously increases in CSGs, reaching $\sim 55\%$ at $z = 0$.

\begin{figure}
    \centering
    \includegraphics[scale=0.5]{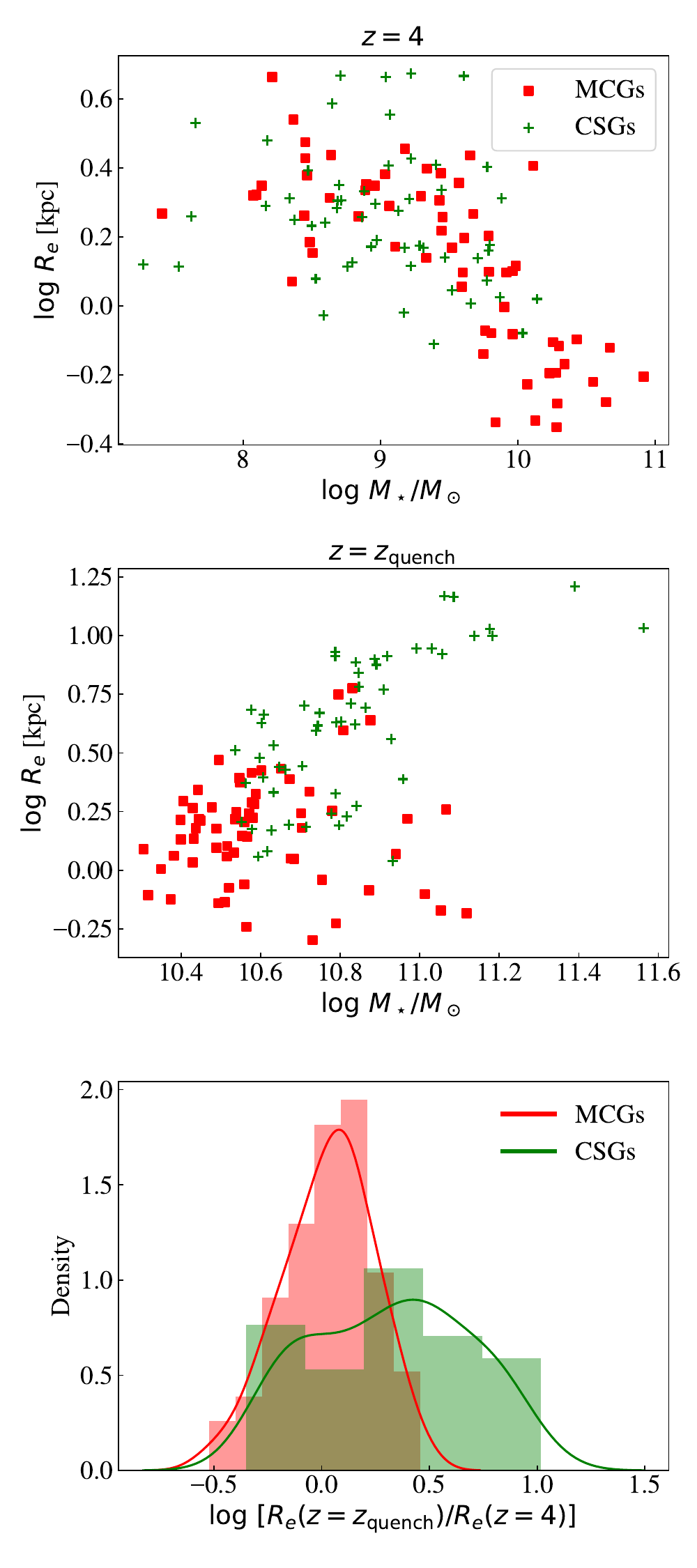}
    \caption{Half-mass radius versus stellar mass for MCGs (red squares) and CSGs (green crosses) progenitors at $z=4$ (top panel) and at the quenching snapshot (middle panel). The bottom panel shows the variation in $\log \, R_e$ between $z=4$ and $z=z_\mathrm{quench}$ with solid lines showing kernel density estimates using Gaussian kernels with adaptive bandwidths following Scott's rule \citep{scottrule}.}
    \label{fig:mass-size_highz}
\end{figure}

\section{Discussion}

\subsection{MCGs in IllustrisTNG and SDSS-MaNGA}

In SM21 we investigated a sample of MCGs in the local universe using data from the MaNGA survey, aiming to characterize the kinematics, morphology and stellar populations of these objects. We found that observed MCGs are predominantly old (but with a tail extending to younger ages), metal rich and $\alpha$-enhanced. Metallicity gradients are slightly negative, while age and $\alpha$-enhancement gradients are close to zero. In terms of kinematics, observed MCGs are rotationally supported, with 96\% being fast rotators. 

Qualitatively IllustrisTNG reproduces most of the observed stellar population properties of MCGs: simulated galaxies are also predominantly old with a tail of younger objects, metal rich and $\alpha$-enhanced. Age and [$\alpha$/Fe] gradients of simulated MCGs are close to zero, as observed. The fraction of fast rotators is slightly smaller in the simulation at 90.5\%, but it still is in reasonable agreement with observations, even more so considering that it has been shown that TNG100 does not reproduce well the morphology of small, more spheroidal, quiescent galaxies due to its limited spatial resolution \citep{Zanisi21,rodrigo22}. The fraction of slow rotators among CSGs (17.5\%), however, is significantly smaller than observed (29.6\%), suggesting that either IllustrisTNG fails to reproduce the kinematics of massive early type galaxies, or that the sizes, masses and velocity dispersion of simulated slow rotators are such that they tare unlikely to satisfy our CSG selection criteria.

\subsection{The Star-Forming Progenitors of MCGs and CSGs}

To assess if MCGs and CSGs can be considered to descend from the same star-forming population, we show in Fig.\,\ref{fig:mass-size_highz} their distribution in the size-mass plane at $z = 4 $ and $z = z_\mathrm{quench}$ together with a histogram of the variation in size between these two epochs. We note that at $z = 4$, despite significant overlap between the distributions, significant differences are already noticeable: the MCG distribution extends to higher stellar masses and smaller sizes. Notably, there is a subset of MCGs with $\log M_\star/M_\odot (z = 4) \gtrsim 10$ and $R_e (z = 4) \lesssim 1$\,kpc with no counterpart among CSGs. At $z = z_\mathrm{quench}$ there is significantly less overlap between distributions. CSGs tend to be more massive than MCGs at the time of quenching. There are no CSGs with $\log M_\star/M_\odot (z = z_\mathrm{quench}) \lesssim 10.6$, and there is a subset of high mass CSGs with no counterparts among MCGs. Moreover, applying our size-mass selection criteria to the CSGs at $z = z_\mathrm{quench}$ we find that 67\% are classified as compact at the time of quenching, so around one third of $z = 0$ massive quiescent galaxies do not have a compact galaxy as its main progenitor, in agreement with previous studies based on the original Illustris run \citep{wellons16}.

\subsection{The Formation Pathways of MCGs}

\citet{du21} studied the growth of galactic structures is TNG50, concluding it can be divided in three regimes: an early-phase evolution at high redshift related to the growth of the bulge, followed by late-phase internal processes and late-phase external interactions which grow the stellar disk and halo respectively. Bulge dominated galaxies such as MCGs follow a compact pathway, which is divided into two phases: I) an early bulge growth phase driven by the accretion of low angular momentum gas, where the galaxy significantly grows in mass while the size changes little; II) a disk growing phase where the size increases significantly with little mass growth.   

Do MCGs fit into this scenario? As seen in Fig.\,\ref{fig:mass-size_highz}, MCGs experience a modest size evolution in general, in line with the description of the first phase of the compact pathway. MCGs do not complete the second phase, however, as they fail to grow a large disk before quenching, unlike CSGs. AGN feedback is responsible for this: the accretion of low angular momentum material favours SMBH growth, consequently the SMBHs in MCGs reach the mass threshold of $\log M_\mathrm{BH} \sim 10^{8.5}\,M_\odot$ - when kinetic AGN feedback is triggered - earlier in their evolution, quenching star formation formation before a large disk is able to growth.

In high-resolution hydrodynamical simulations, compact galaxies are formed in compaction events, when a drastic loss of angular momentum drives large amounts of gas to the center of a galaxy, fueling a nuclear starburst and reducing the size of the galaxy \citep{zolotov15,lapiner23}. The star-formation is quenched soon after due to a combination of gas consumption and feedback \citep{tacchella16}. A small fraction of MCGs experienced a significant decrease in size between $z = 4$  and $z = z_\mathrm{quench}$, 12.7\% experience a decrease of 40\% or more, so wet compaction appears to be a subdominant mechanism for the formation of MCGs. Mergers are a frequent trigger of wet compaction: of the eight MCGs that went through wet compaction, four experienced minor mergers and one experienced a major merger in the 1\,Gyr before quenching.

After quenching, MCGs and CSGs follow very different paths. In general, MCGs have a quiet accretion history, 71.4\% do not merge, compared to 36.5\% of CSGs. At $z = 0$, MCGs preferentially inhabit environments where mergers are less likely; they tend to be either centrals in low mass halos or satellites in galaxy clusters. 

How does the formation scenario described above compares to the scenario proposed by SM21? In the scenario of SM21, MCGs experienced highly-dissipative gas-rich events followed by an intense, short, and centrally concentrated burst of star formation after which the galaxy rapidly quenches and passes through a post-starburst phase. This is very different from the picture painted by IllustrisTNG. While some MCGs do go through a starburst before quenching (see the 84th percentile curve in the top right panel of Fig.\,\ref{fig:bhdot_mgas_sfr}), most keep an approximately constant SFR (and sSFR) in the 1\,Gyr prior to quenching. Once kinetic AGN feedback kicks in, the sSFRs drop very fast, so MCGs are indeed fast-quenched, as proposed by SM21. Highly-dissipative gas-rich events are subdominant in IllustrisTNG, the small size of MCGs are mainly due to an early formation and growth by accretion of low angular momentum gas

\section{Summary and Conclusions}

In this work we analyzed a sample of 63 massive compact galaxies extracted from the TNG100 simulation, comparing their $z = 0$ stellar population properties and kinematics to observations. We also traced these galaxies back in time, assessing their star formation history, size evolution, merger history and supermassive black hole assembly history. We found that:

\begin{itemize}
    \item MCGs are mainly old (median age $\sim$ 10.8\,Gyr), have super-solar metallicities (median $\log Z/Z_{\odot} \sim 0.35$) and $\alpha$-enhanced (median $[\alpha/Fe] \sim 0.25$). The age distribution extends to younger ages, however, and a few MCGs are as young $\sim 7$\,Gyr. Age and $[\alpha/Fe]$ gradients are flat. In terms of stellar kinematics, MCGs are fast-rotators (90.5\%). These properties are in qualitive agreement with observations; 
    \item Compared to a control sample of median sized quiescent galaxies matched in velocity dispersion, MCGs are older and more metal rich. There are no significant differences between the $[\alpha/Fe]$ distributions;
    \item In general, MCGs assemble their mass early and accrete low angular momentum gas, significantly increasing their mass with little size growth. Approximately 13\% of MCGs go through a compaction event, their sizes shrinking by 40\% or more;
    \item The accretion of low angular momentum gas leads to enhanced SMBH growth: MCGs have larger $M_\mathrm{BH}/M_\star$ than CSGs. They also reach the threshold SMBH mass of $\log M_\mathrm{BH} \sim 10^{8.5}\,M_\odot$ - when kinetic AGN feedback kicks in and quenches the galaxy - earlier;
    \item Comparing the size evolution between $z = 4$ and $z = z_\mathrm{quench}$ of MCGs and CSGs we find that 67\% of CSGs are compact at the time of quenching. This implies that a least a third of $z = 0$ massive quiescent galaxies do not have a compact galaxy as their main progenitor;
    \item MCGs have a quiet accretion history: 71\% do not merge after quenching.
\end{itemize}

\section*{Acknowledgements}

The authors thank the anonymous referee for their comments and suggestions, which led to an improved version of the manuscript. This work resulted from the MSc dissertation developed by the author to obtain his degree at the Federal University of Rio Grande do Sul (UFRGS), which was supervised by ASM.
The author thus thanks ASM's supervision as well as the MSc defense committee, composed of Marina Trevisan, Cristina Furlanetto and Rogemar A. Riffel, for their useful comments on this work.
FSL and ASM acknowledge the meaningful discussions with Rodrigo Flores-Freitas that positively impacted our work.
The author also acknowledge the funding agency \textit{Conselho Nacional de Desenvolvimento Científico e Tecnológico} (CNPq) for funding this work through grant number 131090/2020-8.
ASM acknowledges the financial support from the Brazilian national council for scientific and technological development (CNPq).
TVR also acknowledges CNPq for funding through grant number 304584/2022-3.
MT thanks the support of CNPq (process 312541/2021-0).
\section*{Data availability}
The data underlying this article are publicly available at \url{www.tng-project.org/data} \citep{nelson19}. Additional data generated by the analyses in this work are available upon request
to the corresponding author.




\bibliographystyle{mnras}
\bibliography{MCG_TNG} 

\begin{thebibliography}{}
\makeatletter
\relax
\def\mn@urlcharsother{\let\do\@makeother \do\$\do\&\do\#\do\^\do\_\do\%\do\~}
\def\mn@doi{\begingroup\mn@urlcharsother \@ifnextchar [ {\mn@doi@}
  {\mn@doi@[]}}
\def\mn@doi@[#1]#2{\def\@tempa{#1}\ifx\@tempa\@empty \href
  {http://dx.doi.org/#2} {doi:#2}\else \href {http://dx.doi.org/#2} {#1}\fi
  \endgroup}
\def\mn@eprint#1#2{\mn@eprint@#1:#2::\@nil}
\def\mn@eprint@arXiv#1{\href {http://arxiv.org/abs/#1} {{\tt arXiv:#1}}}
\def\mn@eprint@dblp#1{\href {http://dblp.uni-trier.de/rec/bibtex/#1.xml}
  {dblp:#1}}
\def\mn@eprint@#1:#2:#3:#4\@nil{\def\@tempa {#1}\def\@tempb {#2}\def\@tempc
  {#3}\ifx \@tempc \@empty \let \@tempc \@tempb \let \@tempb \@tempa \fi \ifx
  \@tempb \@empty \def\@tempb {arXiv}\fi \@ifundefined
  {mn@eprint@\@tempb}{\@tempb:\@tempc}{\expandafter \expandafter \csname
  mn@eprint@\@tempb\endcsname \expandafter{\@tempc}}}

\bibitem[\protect\citeauthoryear{{Aguado} et~al.,}{{Aguado}
  et~al.}{2019}]{aguado19}
{Aguado} D.~S.,  et~al., 2019, \mn@doi [\apjs]
  {10.3847/1538-4365/aaf65110.48550/arXiv.1812.02759}, \href
  {https://ui.adsabs.harvard.edu/abs/2019ApJS..240...23A} {240, 23}

\bibitem[\protect\citeauthoryear{{Aird}, {Coil}  \& {Kocevski}}{{Aird}
  et~al.}{2022}]{aird22}
{Aird} J.,  {Coil} A.~L.,   {Kocevski} D.~D.,  2022, \mn@doi [\mnras]
  {10.1093/mnras/stac2103}, \href
  {https://ui.adsabs.harvard.edu/abs/2022MNRAS.515.4860A} {515, 4860}

\bibitem[\protect\citeauthoryear{{Athanassoula}}{{Athanassoula}}{1984}]{athanassoula84}
{Athanassoula} E.,  1984, \physrep, \href
  {https://ui.adsabs.harvard.edu/abs/1984PhR...114..321A} {114, 321}

\bibitem[\protect\citeauthoryear{{Barro} et~al.,}{{Barro}
  et~al.}{2013}]{barro13}
{Barro} G.,  et~al., 2013, \mn@doi [\apj] {10.1088/0004-637X/765/2/104}, \href
  {https://ui.adsabs.harvard.edu/abs/2013ApJ...765..104B} {765, 104}

\bibitem[\protect\citeauthoryear{{Barro} et~al.,}{{Barro}
  et~al.}{2017a}]{barro17}
{Barro} G.,  et~al., 2017a, \mn@doi [\apj] {10.3847/1538-4357/aa6b05}, \href
  {https://ui.adsabs.harvard.edu/abs/2017ApJ...840...47B} {840, 47}

\bibitem[\protect\citeauthoryear{{Barro} et~al.,}{{Barro}
  et~al.}{2017b}]{barro17b}
{Barro} G.,  et~al., 2017b, \mn@doi [\apjl] {10.3847/2041-8213/aa9f0d}, \href
  {https://ui.adsabs.harvard.edu/abs/2017ApJ...851L..40B} {851, L40}

\bibitem[\protect\citeauthoryear{{Beasley}, {Trujillo}, {Leaman}  \&
  {Montes}}{{Beasley} et~al.}{2018}]{beasley18}
{Beasley} M.~A.,  {Trujillo} I.,  {Leaman} R.,   {Montes} M.,  2018, \mn@doi
  [\nat] {10.1038/nature25756}, \href
  {https://ui.adsabs.harvard.edu/abs/2018Natur.555..483B} {555, 483}

\bibitem[\protect\citeauthoryear{{Belli}, {Newman}  \& {Ellis}}{{Belli}
  et~al.}{2019}]{belli19}
{Belli} S.,  {Newman} A.~B.,   {Ellis} R.~S.,  2019, \mn@doi [\apj]
  {10.3847/1538-4357/ab07af}, \href
  {https://ui.adsabs.harvard.edu/abs/2019ApJ...874...17B} {874, 17}

\bibitem[\protect\citeauthoryear{{Brinchmann}, {Charlot}, {White}, {Tremonti},
  {Kauffmann}, {Heckman}  \& {Brinkmann}}{{Brinchmann}
  et~al.}{2004}]{brinchmann04}
{Brinchmann} J.,  {Charlot} S.,  {White} S.~D.~M.,  {Tremonti} C.,  {Kauffmann}
  G.,  {Heckman} T.,   {Brinkmann} J.,  2004, \mn@doi [\mnras]
  {10.1111/j.1365-2966.2004.07881.x}, \href
  {https://ui.adsabs.harvard.edu/abs/2004MNRAS.351.1151B} {351, 1151}

\bibitem[\protect\citeauthoryear{{Buitrago} et~al.,}{{Buitrago}
  et~al.}{2018}]{buitrago18}
{Buitrago} F.,  et~al., 2018, \mn@doi [\aap] {10.1051/0004-6361/201833785},
  \href {https://ui.adsabs.harvard.edu/abs/2018A&A...619A.137B} {619, A137}

\bibitem[\protect\citeauthoryear{{Bundy} et~al.,}{{Bundy}
  et~al.}{2015}]{bundy15}
{Bundy} K.,  et~al., 2015, \mn@doi [\apj] {10.1088/0004-637X/798/1/7}, \href
  {https://ui.adsabs.harvard.edu/abs/2015ApJ...798....7B} {798, 7}

\bibitem[\protect\citeauthoryear{{Cappellari}}{{Cappellari}}{2016}]{cappellari16}
{Cappellari} M.,  2016, \mn@doi [\araa] {10.1146/annurev-astro-082214-122432},
  \href {https://ui.adsabs.harvard.edu/abs/2016ARA&A..54..597C} {54, 597}

\bibitem[\protect\citeauthoryear{{Cappellari} et~al.,}{{Cappellari}
  et~al.}{2007}]{cappellari07}
{Cappellari} M.,  et~al., 2007, \mn@doi [\mnras]
  {10.1111/j.1365-2966.2007.11963.x}, \href
  {https://ui.adsabs.harvard.edu/abs/2007MNRAS.379..418C} {379, 418}

\bibitem[\protect\citeauthoryear{{Carnall} et~al.,}{{Carnall}
  et~al.}{2023}]{carnall23}
{Carnall} A.~C.,  et~al., 2023, \mn@doi [\mnras] {10.1093/mnras/stad369}, \href
  {https://ui.adsabs.harvard.edu/abs/2023MNRAS.520.3974C} {520, 3974}

\bibitem[\protect\citeauthoryear{{Carollo} et~al.,}{{Carollo}
  et~al.}{2013}]{carollo13}
{Carollo} C.~M.,  et~al., 2013, \mn@doi [\apj] {10.1088/0004-637X/773/2/112},
  \href {https://ui.adsabs.harvard.edu/abs/2013ApJ...773..112C} {773, 112}

\bibitem[\protect\citeauthoryear{{Conroy}}{{Conroy}}{2013}]{conroy13}
{Conroy} C.,  2013, \mn@doi [\araa] {10.1146/annurev-astro-082812-141017},
  \href {https://ui.adsabs.harvard.edu/abs/2013ARA&A..51..393C} {51, 393}

\bibitem[\protect\citeauthoryear{{Damjanov}, {Zahid}, {Geller}, {Utsumi},
  {Sohn}  \& {Souchereau}}{{Damjanov} et~al.}{2019}]{damjanov19}
{Damjanov} I.,  {Zahid} H.~J.,  {Geller} M.~J.,  {Utsumi} Y.,  {Sohn} J.,
  {Souchereau} H.,  2019, \mn@doi [\apj] {10.3847/1538-4357/aaf97d}, \href
  {https://ui.adsabs.harvard.edu/abs/2019ApJ...872...91D} {872, 91}

\bibitem[\protect\citeauthoryear{{Davies}, {Pontzen}  \& {Crain}}{{Davies}
  et~al.}{2022}]{davies22}
{Davies} J.~J.,  {Pontzen} A.,   {Crain} R.~A.,  2022, \mn@doi [\mnras]
  {10.1093/mnras/stac1742}, \href
  {https://ui.adsabs.harvard.edu/abs/2022MNRAS.515.1430D} {515, 1430}

\bibitem[\protect\citeauthoryear{{Dekel} \& {Burkert}}{{Dekel} \&
  {Burkert}}{2014}]{dekel14}
{Dekel} A.,  {Burkert} A.,  2014, \mn@doi [\mnras] {10.1093/mnras/stt2331},
  \href {https://ui.adsabs.harvard.edu/abs/2014MNRAS.438.1870D} {438, 1870}

\bibitem[\protect\citeauthoryear{{Di Matteo}, {Springel}  \& {Hernquist}}{{Di
  Matteo} et~al.}{2005}]{dimatteo05}
{Di Matteo} T.,  {Springel} V.,   {Hernquist} L.,  2005, \mn@doi [\nat]
  {10.1038/nature03335}, \href
  {https://ui.adsabs.harvard.edu/abs/2005Natur.433..604D} {433, 604}

\bibitem[\protect\citeauthoryear{{Dolag}, {Borgani}, {Murante}  \&
  {Springel}}{{Dolag} et~al.}{2009}]{dolag09}
{Dolag} K.,  {Borgani} S.,  {Murante} G.,   {Springel} V.,  2009, \mn@doi
  [\mnras] {10.1111/j.1365-2966.2009.15034.x}, \href
  {https://ui.adsabs.harvard.edu/abs/2009MNRAS.399..497D} {399, 497}

\bibitem[\protect\citeauthoryear{{Donnari} et~al.}{{Donnari}
  et~al.}{2019}]{donnari19}
{Donnari} M.,  et~al., 2019, \mn@doi [MNRAS] {10.1093/mnras/stz712}, \href
  {https://ui.adsabs.harvard.edu/abs/2019MNRAS.485.4817D} {485, 4817}

\bibitem[\protect\citeauthoryear{{Donnari} et~al.,}{{Donnari}
  et~al.}{2021}]{donnari21}
{Donnari} M.,  et~al., 2021, \mn@doi [\mnras] {10.1093/mnras/staa3006}, \href
  {https://ui.adsabs.harvard.edu/abs/2021MNRAS.500.4004D} {500, 4004}

\bibitem[\protect\citeauthoryear{{Drory} et~al.,}{{Drory}
  et~al.}{2015}]{drory15}
{Drory} N.,  et~al., 2015, \mn@doi [\aj] {10.1088/0004-6256/149/2/77}, \href
  {https://ui.adsabs.harvard.edu/abs/2015AJ....149...77D} {149, 77}

\bibitem[\protect\citeauthoryear{{Du}, {Ho}, {Debattista}, {Pillepich},
  {Nelson}, {Hernquist}  \& {Weinberger}}{{Du} et~al.}{2021}]{du21}
{Du} M.,  {Ho} L.~C.,  {Debattista} V.~P.,  {Pillepich} A.,  {Nelson} D.,
  {Hernquist} L.,   {Weinberger} R.,  2021, \mn@doi [\apj]
  {10.3847/1538-4357/ac0e98}, \href
  {https://ui.adsabs.harvard.edu/abs/2021ApJ...919..135D} {919, 135}

\bibitem[\protect\citeauthoryear{{Ellison} et~al.,}{{Ellison}
  et~al.}{2022}]{ellison22}
{Ellison} S.~L.,  et~al., 2022, \mn@doi [\mnras] {10.1093/mnrasl/slac109},
  \href {https://ui.adsabs.harvard.edu/abs/2022MNRAS.517L..92E} {517, L92}

\bibitem[\protect\citeauthoryear{{Emsellem} et~al.,}{{Emsellem}
  et~al.}{2007}]{emsellem07}
{Emsellem} E.,  et~al., 2007, \mn@doi [\mnras]
  {10.1111/j.1365-2966.2007.11752.x}, \href
  {https://ui.adsabs.harvard.edu/abs/2007MNRAS.379..401E} {379, 401}

\bibitem[\protect\citeauthoryear{{Fagioli} et~al.}{{Fagioli}
  et~al.}{2016}]{fagioli16}
{Fagioli} M.,  et~al., 2016, \mn@doi [ApJ] {10.3847/0004-637X/831/2/173}, \href
  {https://ui.adsabs.harvard.edu/abs/2016ApJ...831..173F} {831, 173}

\bibitem[\protect\citeauthoryear{{Faisst}, {Carollo}, {Capak}, {Tacchella},
  {Renzini}, {Ilbert}, {McCracken}  \& {Scoville}}{{Faisst}
  et~al.}{2017}]{faisst17}
{Faisst} A.~L.,  {Carollo} C.~M.,  {Capak} P.~L.,  {Tacchella} S.,  {Renzini}
  A.,  {Ilbert} O.,  {McCracken} H.~J.,   {Scoville} N.~Z.,  2017, \mn@doi
  [\apj] {10.3847/1538-4357/aa697a}, \href
  {https://ui.adsabs.harvard.edu/abs/2017ApJ...839...71F} {839, 71}

\bibitem[\protect\citeauthoryear{{Ferr{\'e}-Mateu}, {Vazdekis}, {Trujillo},
  {S{\'a}nchez-Bl{\'a}zquez}, {Ricciardelli}  \& {de la
  Rosa}}{{Ferr{\'e}-Mateu} et~al.}{2012}]{ferre12}
{Ferr{\'e}-Mateu} A.,  {Vazdekis} A.,  {Trujillo} I.,
  {S{\'a}nchez-Bl{\'a}zquez} P.,  {Ricciardelli} E.,   {de la Rosa} I.~G.,
  2012, \mn@doi [\mnras] {10.1111/j.1365-2966.2012.20897.x}, \href
  {https://ui.adsabs.harvard.edu/abs/2012MNRAS.423..632F} {423, 632}

\bibitem[\protect\citeauthoryear{{Flores-Freitas}, {Chies-Santos},
  {Furlanetto}, {De Rossi}, {Ferreira}, {Zenocratti}  \&
  {Alamo-Mart{\'\i}nez}}{{Flores-Freitas} et~al.}{2022}]{rodrigo22}
{Flores-Freitas} R.,  {Chies-Santos} A.~L.,  {Furlanetto} C.,  {De Rossi}
  M.~E.,  {Ferreira} L.,  {Zenocratti} L.~J.,   {Alamo-Mart{\'\i}nez} K.~A.,
  2022, \mn@doi [\mnras] {10.1093/mnras/stac187}, \href
  {https://ui.adsabs.harvard.edu/abs/2022MNRAS.512..245F} {512, 245}

\bibitem[\protect\citeauthoryear{{Furlong} et~al.,}{{Furlong}
  et~al.}{2015}]{furlong15}
{Furlong} M.,  et~al., 2015, \mn@doi [\mnras] {10.1093/mnras/stv852}, \href
  {https://ui.adsabs.harvard.edu/abs/2015MNRAS.450.4486F} {450, 4486}

\bibitem[\protect\citeauthoryear{{Genel} et~al.,}{{Genel}
  et~al.}{2014}]{genel14}
{Genel} S.,  et~al., 2014, \mn@doi [\mnras] {10.1093/mnras/stu1654}, \href
  {https://ui.adsabs.harvard.edu/abs/2014MNRAS.445..175G} {445, 175}

\bibitem[\protect\citeauthoryear{{Genel} et~al.,}{{Genel}
  et~al.}{2018}]{genel18}
{Genel} S.,  et~al., 2018, \mn@doi [\mnras]
  {10.1093/mnras/stx307810.48550/arXiv.1707.05327}, \href
  {https://ui.adsabs.harvard.edu/abs/2018MNRAS.474.3976G} {474, 3976}

\bibitem[\protect\citeauthoryear{{Greene}, {Murphy}, {Graves}, {Gunn},
  {Raskutti}, {Comerford}  \& {Gebhardt}}{{Greene} et~al.}{2013}]{greene13}
{Greene} J.~E.,  {Murphy} J.~D.,  {Graves} G.~J.,  {Gunn} J.~E.,  {Raskutti}
  S.,  {Comerford} J.~M.,   {Gebhardt} K.,  2013, \mn@doi [\apj]
  {10.1088/0004-637X/776/2/64}, \href
  {https://ui.adsabs.harvard.edu/abs/2013ApJ...776...64G} {776, 64}

\bibitem[\protect\citeauthoryear{{Hopkins}, {Hernquist}, {Cox}  \&
  {Kere{\v{s}}}}{{Hopkins} et~al.}{2008}]{hopkins08}
{Hopkins} P.~F.,  {Hernquist} L.,  {Cox} T.~J.,   {Kere{\v{s}}} D.,  2008,
  \mn@doi [\apjs] {10.1086/524362}, \href
  {https://ui.adsabs.harvard.edu/abs/2008ApJS..175..356H} {175, 356}

\bibitem[\protect\citeauthoryear{{Huang} et~al.}{{Huang}
  et~al.}{2013}]{huang13}
{Huang} S.,  et~al., 2013, \mn@doi [ApJ] {10.1088/0004-637X/766/1/47}, \href
  {https://ui.adsabs.harvard.edu/abs/2013ApJ...766...47H} {766, 47}

\bibitem[\protect\citeauthoryear{{Ilbert} et~al.,}{{Ilbert}
  et~al.}{2013}]{Ilbert13}
{Ilbert} O.,  et~al., 2013, \mn@doi [\aap] {10.1051/0004-6361/201321100}, \href
  {https://ui.adsabs.harvard.edu/abs/2013A&A...556A..55I} {556, A55}

\bibitem[\protect\citeauthoryear{{Kocevski} et~al.,}{{Kocevski}
  et~al.}{2017}]{kocevski17}
{Kocevski} D.~D.,  et~al., 2017, \mn@doi [\apj] {10.3847/1538-4357/aa8566},
  \href {https://ui.adsabs.harvard.edu/abs/2017ApJ...846..112K} {846, 112}

\bibitem[\protect\citeauthoryear{{Lapiner} et~al.,}{{Lapiner}
  et~al.}{2023}]{lapiner23}
{Lapiner} S.,  et~al., 2023, \mn@doi [\mnras] {10.1093/mnras/stad1263}, \href
  {https://ui.adsabs.harvard.edu/abs/2023MNRAS.tmp.1180L} {}

\bibitem[\protect\citeauthoryear{{Lovell} et~al.}{{Lovell}
  et~al.}{2018}]{lovell18}
{Lovell} M.~R.,  et~al., 2018, \mn@doi [MNRAS] {10.1093/mnras/sty2339}, \href
  {https://ui.adsabs.harvard.edu/abs/2018MNRAS.481.1950L} {481, 1950}

\bibitem[\protect\citeauthoryear{{Lu} et~al.}{{Lu} et~al.}{2020a}]{lu20}
{Lu} S.,  et~al., 2020a, \mn@doi [MNRAS] {10.1093/mnras/staa173}, \href
  {https://ui.adsabs.harvard.edu/abs/2020MNRAS.492.5930L} {492, 5930}

\bibitem[\protect\citeauthoryear{{Lu}, {Cappellari}, {Mao}, {Ge}  \& {Li}}{{Lu}
  et~al.}{2020b}]{lu_cappellari20}
{Lu} S.,  {Cappellari} M.,  {Mao} S.,  {Ge} J.,   {Li} R.,  2020b, \mn@doi
  [\mnras] {10.1093/mnras/staa1481}, \href
  {https://ui.adsabs.harvard.edu/abs/2020MNRAS.495.4820L} {495, 4820}

\bibitem[\protect\citeauthoryear{{Marinacci} et~al.,}{{Marinacci}
  et~al.}{2018}]{marinacci18}
{Marinacci} F.,  et~al., 2018, \mn@doi [\mnras] {10.1093/mnras/sty2206}, \href
  {https://ui.adsabs.harvard.edu/abs/2018MNRAS.480.5113M} {480, 5113}

\bibitem[\protect\citeauthoryear{{Muzzin} et~al.}{{Muzzin}
  et~al.}{2013}]{muzzin13}
{Muzzin} A.,  et~al., 2013, \mn@doi [ApJ] {10.1088/0004-637X/777/1/18}, \href
  {https://ui.adsabs.harvard.edu/abs/2013ApJ...777...18M} {777, 18}

\bibitem[\protect\citeauthoryear{{Naiman} et~al.,}{{Naiman}
  et~al.}{2018}]{naiman18}
{Naiman} J.~P.,  et~al., 2018, \mn@doi [\mnras] {10.1093/mnras/sty618}, \href
  {https://ui.adsabs.harvard.edu/abs/2018MNRAS.477.1206N} {477, 1206}

\bibitem[\protect\citeauthoryear{{Nelson} et~al.,}{{Nelson}
  et~al.}{2015}]{nelson15}
{Nelson} D.,  et~al., 2015, \mn@doi [Astronomy and Computing]
  {10.1016/j.ascom.2015.09.003}, \href
  {https://ui.adsabs.harvard.edu/abs/2015A&C....13...12N} {13, 12}

\bibitem[\protect\citeauthoryear{{Nelson} et~al.,}{{Nelson}
  et~al.}{2018}]{nelson18}
{Nelson} D.,  et~al., 2018, \mn@doi [\mnras] {10.1093/mnras/stx3040}, \href
  {https://ui.adsabs.harvard.edu/abs/2018MNRAS.475..624N} {475, 624}

\bibitem[\protect\citeauthoryear{{Nelson} et~al.,}{{Nelson}
  et~al.}{2019}]{nelson19}
{Nelson} D.,  et~al., 2019, \mn@doi [Computational Astrophysics and Cosmology]
  {10.1186/s40668-019-0028-x}, \href
  {https://ui.adsabs.harvard.edu/abs/2019ComAC...6....2N} {6, 2}

\bibitem[\protect\citeauthoryear{{Noeske} et~al.,}{{Noeske}
  et~al.}{2007}]{noeske07}
{Noeske} K.~G.,  et~al., 2007, \mn@doi [\apjl] {10.1086/517926}, \href
  {https://ui.adsabs.harvard.edu/abs/2007ApJ...660L..43N} {660, L43}

\bibitem[\protect\citeauthoryear{{Oh}, {Greene}  \& {Lackner}}{{Oh}
  et~al.}{2017}]{oh17}
{Oh} S.,  {Greene} J.~E.,   {Lackner} C.~N.,  2017, \mn@doi [\apj]
  {10.3847/1538-4357/836/1/115}, \href
  {https://ui.adsabs.harvard.edu/abs/2017ApJ...836..115O} {836, 115}

\bibitem[\protect\citeauthoryear{{Oser} et~al.}{{Oser} et~al.}{2010}]{oser10}
{Oser} L.,  et~al., 2010, \mn@doi [ApJ] {10.1088/0004-637X/725/2/2312}, \href
  {https://ui.adsabs.harvard.edu/abs/2010ApJ...725.2312O} {725, 2312}

\bibitem[\protect\citeauthoryear{{Pillepich} et~al.,}{{Pillepich}
  et~al.}{2018a}]{pillepich18b}
{Pillepich} A.,  et~al., 2018a, \mn@doi [\mnras] {10.1093/mnras/stx2656}, \href
  {https://ui.adsabs.harvard.edu/abs/2018MNRAS.473.4077P} {473, 4077}

\bibitem[\protect\citeauthoryear{{Pillepich} et~al.,}{{Pillepich}
  et~al.}{2018b}]{pillepich18a}
{Pillepich} A.,  et~al., 2018b, \mn@doi [\mnras] {10.1093/mnras/stx3112}, \href
  {https://ui.adsabs.harvard.edu/abs/2018MNRAS.475..648P} {475, 648}

\bibitem[\protect\citeauthoryear{{Planck Collaboration} et~al.,}{{Planck
  Collaboration} et~al.}{2016}]{planck16}
{Planck Collaboration} et~al., 2016, \mn@doi [\aap]
  {10.1051/0004-6361/201525830}, \href
  {https://ui.adsabs.harvard.edu/abs/2016A&A...594A..13P} {594, A13}

\bibitem[\protect\citeauthoryear{{Pontzen}, {Tremmel}, {Roth}, {Peiris},
  {Saintonge}, {Volonteri}, {Quinn}  \& {Governato}}{{Pontzen}
  et~al.}{2017}]{pontzen17}
{Pontzen} A.,  {Tremmel} M.,  {Roth} N.,  {Peiris} H.~V.,  {Saintonge} A.,
  {Volonteri} M.,  {Quinn} T.,   {Governato} F.,  2017, \mn@doi [\mnras]
  {10.1093/mnras/stw2627}, \href
  {https://ui.adsabs.harvard.edu/abs/2017MNRAS.465..547P} {465, 547}

\bibitem[\protect\citeauthoryear{{Quai}, {Hani}, {Ellison}, {Patton}  \&
  {Woo}}{{Quai} et~al.}{2021}]{quai21}
{Quai} S.,  {Hani} M.~H.,  {Ellison} S.~L.,  {Patton} D.~R.,   {Woo} J.,  2021,
  \mn@doi [\mnras] {10.1093/mnras/stab988}, \href
  {https://ui.adsabs.harvard.edu/abs/2021MNRAS.504.1888Q} {504, 1888}

\bibitem[\protect\citeauthoryear{{Rodriguez-Gomez} et~al.,}{{Rodriguez-Gomez}
  et~al.}{2015}]{RG15}
{Rodriguez-Gomez} V.,  et~al., 2015, \mn@doi [\mnras] {10.1093/mnras/stv264},
  \href {https://ui.adsabs.harvard.edu/abs/2015MNRAS.449...49R} {449, 49}

\bibitem[\protect\citeauthoryear{{Rodriguez-Gomez} et~al.,}{{Rodriguez-Gomez}
  et~al.}{2016}]{rodriguez-gomez16}
{Rodriguez-Gomez} V.,  et~al., 2016, \mn@doi [\mnras] {10.1093/mnras/stw456},
  \href {https://ui.adsabs.harvard.edu/abs/2016MNRAS.458.2371R} {458, 2371}

\bibitem[\protect\citeauthoryear{{Rodriguez-Gomez}, {Snyder}
  et~al.}{{Rodriguez-Gomez} et~al.}{2019}]{rodriguez-gomez19}
{Rodriguez-Gomez} V.,  {Snyder} G.~F.,   et~al., 2019, \mn@doi [MNRAS]
  {10.1093/mnras/sty3345}, \href
  {https://ui.adsabs.harvard.edu/abs/2019MNRAS.483.4140R} {483, 4140}

\bibitem[\protect\citeauthoryear{Rosenbaum \& Rubin}{Rosenbaum \&
  Rubin}{1983}]{psm}
Rosenbaum P.~R.,  Rubin D.~B.,  1983, \mn@doi [Biometrika]
  {10.1093/biomet/70.1.41}, 70, 41

\bibitem[\protect\citeauthoryear{{Sazonova} et~al.,}{{Sazonova}
  et~al.}{2021}]{sazonova21}
{Sazonova} E.,  et~al., 2021, \mn@doi [\apj] {10.3847/1538-4357/ac0f7f}, \href
  {https://ui.adsabs.harvard.edu/abs/2021ApJ...919..134S} {919, 134}

\bibitem[\protect\citeauthoryear{{Schnorr-M{\"u}ller}
  et~al.,}{{Schnorr-M{\"u}ller} et~al.}{2021}]{papermanga}
{Schnorr-M{\"u}ller} A.,  et~al., 2021, \mn@doi [\mnras]
  {10.1093/mnras/stab2116}, \href
  {https://ui.adsabs.harvard.edu/abs/2021MNRAS.507..300S} {507, 300}

\bibitem[\protect\citeauthoryear{Scott}{Scott}{1992}]{scottrule}
Scott D.,  1992, Multivariate Density Estimation: Theory, Practice, and
  Visualization.
A Wiley-interscience publication, Wiley, \url
  {https://books.google.com.br/books?id=7crCUS\_F2ocC}

\bibitem[\protect\citeauthoryear{{Smee} et~al.,}{{Smee} et~al.}{2013}]{smee13}
{Smee} S.~A.,  et~al., 2013, \mn@doi [\aj] {10.1088/0004-6256/146/2/32}, \href
  {https://ui.adsabs.harvard.edu/abs/2013AJ....146...32S} {146, 32}

\bibitem[\protect\citeauthoryear{{Spilker}, {Bezanson}, {Marrone}, {Weiner},
  {Whitaker}  \& {Williams}}{{Spilker} et~al.}{2016}]{spilker16}
{Spilker} J.~S.,  {Bezanson} R.,  {Marrone} D.~P.,  {Weiner} B.~J.,  {Whitaker}
  K.~E.,   {Williams} C.~C.,  2016, \mn@doi [\apj]
  {10.3847/0004-637X/832/1/19}, \href
  {https://ui.adsabs.harvard.edu/abs/2016ApJ...832...19S} {832, 19}

\bibitem[\protect\citeauthoryear{{Spilker}, {Bezanson}, {Weiner}, {Whitaker}
  \& {Williams}}{{Spilker} et~al.}{2019}]{spilker19}
{Spilker} J.~S.,  {Bezanson} R.,  {Weiner} B.~J.,  {Whitaker} K.~E.,
  {Williams} C.~C.,  2019, \mn@doi [\apj] {10.3847/1538-4357/ab3804}, \href
  {https://ui.adsabs.harvard.edu/abs/2019ApJ...883...81S} {883, 81}

\bibitem[\protect\citeauthoryear{{Spiniello} et~al.,}{{Spiniello}
  et~al.}{2021}]{spiniello21}
{Spiniello} C.,  et~al., 2021, \mn@doi [\aap] {10.1051/0004-6361/202038936},
  \href {https://ui.adsabs.harvard.edu/abs/2021A&A...646A..28S} {646, A28}

\bibitem[\protect\citeauthoryear{{Springel}}{{Springel}}{2010}]{springel10}
{Springel} V.,  2010, \mn@doi [\mnras] {10.1111/j.1365-2966.2009.15715.x},
  \href {https://ui.adsabs.harvard.edu/abs/2010MNRAS.401..791S} {401, 791}

\bibitem[\protect\citeauthoryear{{Springel}, {White}, {Tormen}  \&
  {Kauffmann}}{{Springel} et~al.}{2001}]{springel01}
{Springel} V.,  {White} S. D.~M.,  {Tormen} G.,   {Kauffmann} G.,  2001,
  \mn@doi [\mnras] {10.1046/j.1365-8711.2001.04912.x}, \href
  {https://ui.adsabs.harvard.edu/abs/2001MNRAS.328..726S} {328, 726}

\bibitem[\protect\citeauthoryear{{Springel} et~al.,}{{Springel}
  et~al.}{2018}]{springel18}
{Springel} V.,  et~al., 2018, \mn@doi [\mnras] {10.1093/mnras/stx3304}, \href
  {https://ui.adsabs.harvard.edu/abs/2018MNRAS.475..676S} {475, 676}

\bibitem[\protect\citeauthoryear{{Straatman} et~al.,}{{Straatman}
  et~al.}{2014}]{straatman14}
{Straatman} C. M.~S.,  et~al., 2014, \mn@doi [\apjl]
  {10.1088/2041-8205/783/1/L14}, \href
  {https://ui.adsabs.harvard.edu/abs/2014ApJ...783L..14S} {783, L14}

\bibitem[\protect\citeauthoryear{{Suess}, {Kriek}, {Price}  \& {Barro}}{{Suess}
  et~al.}{2021}]{suess21}
{Suess} K.~A.,  {Kriek} M.,  {Price} S.~H.,   {Barro} G.,  2021, \mn@doi [\apj]
  {10.3847/1538-4357/abf1e4}, \href
  {https://ui.adsabs.harvard.edu/abs/2021ApJ...915...87S} {915, 87}

\bibitem[\protect\citeauthoryear{{Tacchella}, {Dekel}, {Carollo}, {Ceverino},
  {DeGraf}, {Lapiner}, {Mandelker}  \& {Primack}}{{Tacchella}
  et~al.}{2016}]{tacchella16}
{Tacchella} S.,  {Dekel} A.,  {Carollo} C.~M.,  {Ceverino} D.,  {DeGraf} C.,
  {Lapiner} S.,  {Mandelker} N.,   {Primack} J.~R.,  2016, \mn@doi [\mnras]
  {10.1093/mnras/stw303}, \href
  {https://ui.adsabs.harvard.edu/abs/2016MNRAS.458..242T} {458, 242}

\bibitem[\protect\citeauthoryear{{Talia} et~al.,}{{Talia}
  et~al.}{2018}]{talia18}
{Talia} M.,  et~al., 2018, \mn@doi [\mnras] {10.1093/mnras/sty481}, \href
  {https://ui.adsabs.harvard.edu/abs/2018MNRAS.476.3956T} {476, 3956}

\bibitem[\protect\citeauthoryear{{Terrazas} et~al.,}{{Terrazas}
  et~al.}{2020}]{terrazas20}
{Terrazas} B.~A.,  et~al., 2020, \mn@doi [\mnras] {10.1093/mnras/staa374},
  \href {https://ui.adsabs.harvard.edu/abs/2020MNRAS.493.1888T} {493, 1888}

\bibitem[\protect\citeauthoryear{{Torrey} et~al.}{{Torrey}
  et~al.}{2019}]{torrey19}
{Torrey} P.,  et~al., 2019, \mn@doi [MNRAS] {10.1093/mnras/stz243}, \href
  {https://ui.adsabs.harvard.edu/abs/2019MNRAS.484.5587T} {484, 5587}

\bibitem[\protect\citeauthoryear{{Verley} et~al.,}{{Verley}
  et~al.}{2007}]{verley07}
{Verley} S.,  et~al., 2007, \mn@doi [\aap] {10.1051/0004-6361:20077307}, \href
  {https://ui.adsabs.harvard.edu/abs/2007A&A...470..505V} {470, 505}

\bibitem[\protect\citeauthoryear{{Verrico} et~al.,}{{Verrico}
  et~al.}{2022}]{verrico22}
{Verrico} M.,  et~al., 2022, \mn@doi [arXiv e-prints]
  {10.48550/arXiv.2211.16532}, \href
  {https://ui.adsabs.harvard.edu/abs/2022arXiv221116532V} {p. arXiv:2211.16532}

\bibitem[\protect\citeauthoryear{{Vogelsberger}, {Genel}, {Sijacki}, {Torrey},
  {Springel}  \& {Hernquist}}{{Vogelsberger} et~al.}{2013}]{vogelsberger13}
{Vogelsberger} M.,  {Genel} S.,  {Sijacki} D.,  {Torrey} P.,  {Springel} V.,
  {Hernquist} L.,  2013, \mn@doi [\mnras] {10.1093/mnras/stt1789}, \href
  {https://ui.adsabs.harvard.edu/abs/2013MNRAS.436.3031V} {436, 3031}

\bibitem[\protect\citeauthoryear{{Vogelsberger} et~al.,}{{Vogelsberger}
  et~al.}{2014a}]{vogelsberger14a}
{Vogelsberger} M.,  et~al., 2014a, \mn@doi [\mnras] {10.1093/mnras/stu1536},
  \href {https://ui.adsabs.harvard.edu/abs/2014MNRAS.444.1518V} {444, 1518}

\bibitem[\protect\citeauthoryear{{Vogelsberger} et~al.,}{{Vogelsberger}
  et~al.}{2014b}]{vogelsberger14b}
{Vogelsberger} M.,  et~al., 2014b, \mn@doi [\nat] {10.1038/nature13316}, \href
  {https://ui.adsabs.harvard.edu/abs/2014Natur.509..177V} {509, 177}

\bibitem[\protect\citeauthoryear{{Walters}, {Woo}, {Ellison}  \&
  {Hani}}{{Walters} et~al.}{2021}]{walters21}
{Walters} D.,  {Woo} J.,  {Ellison} S.~L.,   {Hani} M.~H.,  2021, \mn@doi
  [\mnras] {10.1093/mnras/stab840}, \href
  {https://ui.adsabs.harvard.edu/abs/2021MNRAS.504.1677W} {504, 1677}

\bibitem[\protect\citeauthoryear{{Wang} et~al.,}{{Wang} et~al.}{2020}]{wang20}
{Wang} Y.,  et~al., 2020, \mn@doi [\mnras] {10.1093/mnras/stz3348}, \href
  {https://ui.adsabs.harvard.edu/abs/2020MNRAS.491.5188W} {491, 5188}

\bibitem[\protect\citeauthoryear{{Weinberger} et~al.,}{{Weinberger}
  et~al.}{2017}]{weinberger17}
{Weinberger} R.,  et~al., 2017, \mn@doi [\mnras] {10.1093/mnras/stw2944}, \href
  {https://ui.adsabs.harvard.edu/abs/2017MNRAS.465.3291W} {465, 3291}

\bibitem[\protect\citeauthoryear{{Weinberger} et~al.}{{Weinberger}
  et~al.}{2018}]{weinberger18}
{Weinberger} R.,  et~al., 2018, \mn@doi [MNRAS] {10.1093/mnras/sty1733}, \href
  {https://ui.adsabs.harvard.edu/abs/2018MNRAS.479.4056W} {479, 4056}

\bibitem[\protect\citeauthoryear{{Wellons} et~al.,}{{Wellons}
  et~al.}{2016}]{wellons16}
{Wellons} S.,  et~al., 2016, \mn@doi [\mnras] {10.1093/mnras/stv2738}, \href
  {https://ui.adsabs.harvard.edu/abs/2016MNRAS.456.1030W} {456, 1030}

\bibitem[\protect\citeauthoryear{{Y{\i}ld{\i}r{\i}m}, {van den Bosch}, {van de
  Ven}, {Mart{\'\i}n-Navarro}, {Walsh}, {Husemann}, {G{\"u}ltekin}  \&
  {Gebhardt}}{{Y{\i}ld{\i}r{\i}m} et~al.}{2017}]{yildirim17}
{Y{\i}ld{\i}r{\i}m} A.,  {van den Bosch} R. C.~E.,  {van de Ven} G.,
  {Mart{\'\i}n-Navarro} I.,  {Walsh} J.~L.,  {Husemann} B.,  {G{\"u}ltekin} K.,
    {Gebhardt} K.,  2017, \mn@doi [\mnras] {10.1093/mnras/stx732}, \href
  {https://ui.adsabs.harvard.edu/abs/2017MNRAS.468.4216Y} {468, 4216}

\bibitem[\protect\citeauthoryear{{Zanisi} et~al.,}{{Zanisi}
  et~al.}{2021}]{Zanisi21}
{Zanisi} L.,  et~al., 2021, \mn@doi [\mnras] {10.1093/mnras/staa3864}, \href
  {https://ui.adsabs.harvard.edu/abs/2021MNRAS.501.4359Z} {501, 4359}

\bibitem[\protect\citeauthoryear{{Zinger} et~al.,}{{Zinger}
  et~al.}{2020}]{zinger20}
{Zinger} E.,  et~al., 2020, \mn@doi [\mnras] {10.1093/mnras/staa2607}, \href
  {https://ui.adsabs.harvard.edu/abs/2020MNRAS.499..768Z} {499, 768}

\bibitem[\protect\citeauthoryear{{Zolotov} et~al.,}{{Zolotov}
  et~al.}{2015}]{zolotov15}
{Zolotov} A.,  et~al., 2015, \mn@doi [\mnras] {10.1093/mnras/stv740}, \href
  {https://ui.adsabs.harvard.edu/abs/2015MNRAS.450.2327Z} {450, 2327}

\bibitem[\protect\citeauthoryear{{van Dokkum} et~al.,}{{van Dokkum}
  et~al.}{2015}]{vandokkum15}
{van Dokkum} P.~G.,  et~al., 2015, \mn@doi [\apj] {10.1088/0004-637X/813/1/23},
  \href {https://ui.adsabs.harvard.edu/abs/2015ApJ...813...23V} {813, 23}

\bibitem[\protect\citeauthoryear{{van der Wel} et~al.}{{van der Wel}
  et~al.}{2014}]{vanderwel14}
{van der Wel} A.,  et~al., 2014, \mn@doi [ApJ] {10.1088/0004-637X/788/1/28},
  \href {https://ui.adsabs.harvard.edu/abs/2014ApJ...788...28V} {788, 28}

\makeatother
\end{thebibliography}




\appendix

\section{Massive Compact Galaxies in TNG50}
\label{appendix_a}

We analysed compact galaxies in TNG50 employing the same methodology described in Section \ref{sec:methodology} in order to investigate how the resolution limits of TNG100 may impact our results, in particular for compact galaxies whose half-mass radii can be only slightly larger than the simulation softening length.
The sample selection resulted in a sample with 9 galaxies, although one galaxy could not be matched in $\sigma_e$ with the control sample and had to be removed from the sample. 
Still, the sample selection in TNG50 resulted in a number density of compact galaxies consistent with TNG100.

The kinematic maps obtained in TNG50 are more detailed in comparison to TNG100, as expected. 
In MCGs, the same overall trend of TNG100 is observed on the maps: velocity maps show a rotation pattern with a rather low peak velocity, and velocity dispersion maps peak in the central region.
The $\lambda_e$-$\epsilon$ diagram in TNG50 reveal that CSGs are fast rotators, while MCGs have significantly lower $\lambda_e$ in comparison, although only one MCG is considered a slow rotator.


The environment quantities $Q_\mathrm{group}$ and $M_\mathrm{halo}$ follow similar trends in TNG50 as in TNG100.
MCGs are found mostly in locally isolated environments with $Q_\mathrm{group} \leq -2$, while most CSGs have $Q_\mathrm{group} > -2$, as is observed in TNG100.
Regarding the global environment, both MCGs and CSGs have a similar distributions of $M_\mathrm{halo}$. 
Unlike TNG100, no bimodal distribution is present for MCGs in TNG50, which can be attributed to the low statistic significance of the TNG50 sample. 

\begin{figure*}
    \centering
    \includegraphics[width=\textwidth]{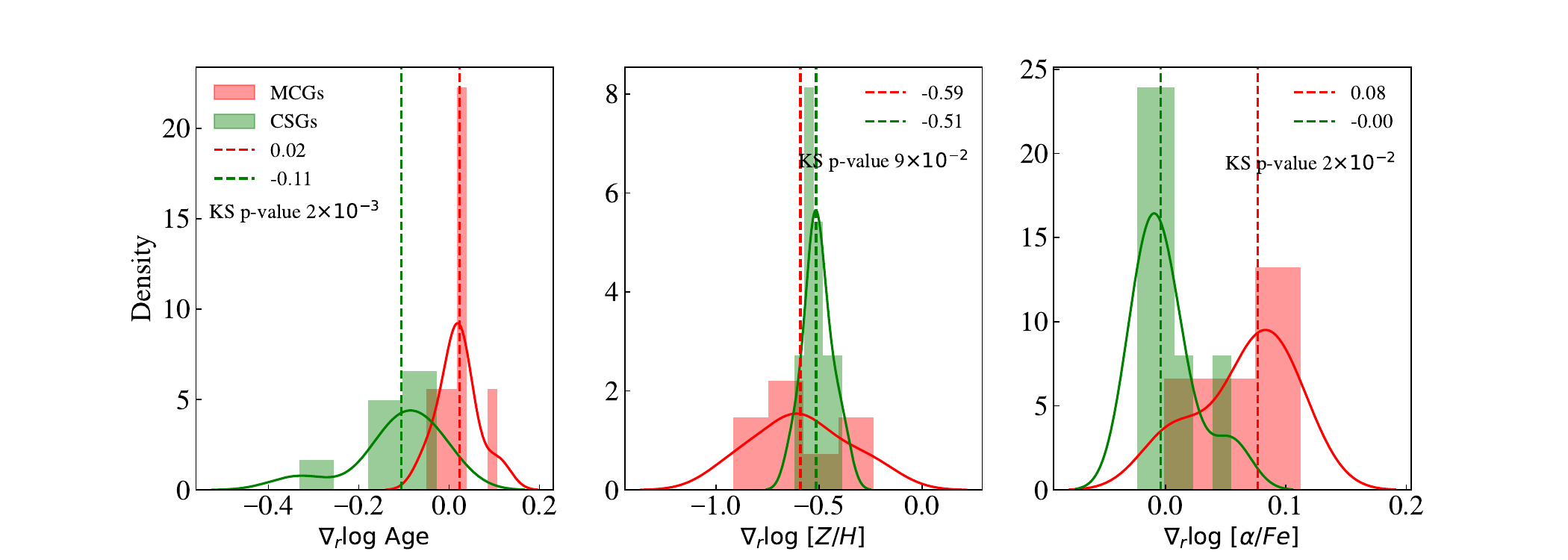}
    \caption{Stellar population gradients in TNG50. From left to right: age gradients, metallicity gradients and [$\alpha$/Fe] gradients. MCGs are shown in red and CSGs are shown in green. Dashed lines indicate the medians of the distributions, and solid lines represent kernel density estimates using Gaussian kernels with adaptive bandwidths following Scott's rule \citep{scottrule}.}
    \label{fig:grads_tng50}
\end{figure*}

The stellar population properties inside $1 R_e$ of MCGs found in TNG100 are all well reproduced in TNG50.
MCGs are older than CSGs, with a median age of $9.6\, \mathrm{Gyr}$ for MCGs and $8.6\, \mathrm{Gyr}$ for CSGs.
MCGs are also more metal rich than CSGs, and both samples have similar $\alpha$ abundances around 0.22.
The respective gradients, however, present a few discrepancies in comparison to TNG100 and are shown in Fig. \ref{fig:grads_tng50}. 
As in TNG100, the gradients were calculated using the inner and outer regions of $R < 1R_e$  and $1R_e < R < 2R_e$, respectively, and employing equations \ref{gradage} and \ref{gradmetal}.
The resulting age gradients are consistent with TNG100, with MCGs having flat age profiles while CSGs have mostly negative gradients.
The metallicity gradients of MCGs, however, are more negative in TNG50, and their [$\alpha$/Fe] gradients are slightly more positive than in TNG100.
The metallicity and [$\alpha$/Fe] gradients of CSGs seem to be consistent with TNG100.


The mass assembly of both samples in TNG50 is qualitatively in agreement with TNG100.
MCGs form their mass early, while CSGs have a more extended formation period, with lookback formation times $(t_{50}, t_{80})_\mathrm{MCGs} = (10.0, 8.2)\, \mathrm{Gyr}$ and $(t_{50}, t_{80})_\mathrm{CSGs} = (8.7, 6.2)\, \mathrm{Gyr}$. These numbers are shifted in $\sim 1 \, \mathrm{Gyr}$ to lower values when compared to TNG100, but convey the same general result.


The supermassive black holes of MCGs and CSGs behave similarly in TNG50 and TNG100. 
In TNG50, they are more massive in CSGs since a lookback time of $\sim 7.5 \, \mathrm{Gyr}$ and are proportionally more massive in MCGs throughout the simulation, in agreement with TNG100.
Although more mildly, the sudden increase in the SMBH mass shortly before quenching can also be seen in TNG50, as well as the peak in  accretion rate and star formation rate.
After quenching, MCGs have no residual star formation rate and no cold gas left, while for CSGs there is still some star formation and gas up to $z=0$, consistent with our results from TNG100.

\bsp	
\label{lastpage}
\end{document}